\documentclass[a4paper,11pt]{article}
\usepackage{booktabs}
\usepackage{bm}
\usepackage{float}
\usepackage{latexsym}
\usepackage{dcolumn}
\usepackage{amsfonts,amssymb}
\usepackage{graphicx}
\usepackage{epsfig}
\usepackage{psfrag}
\usepackage{nccmath}
\usepackage{moresize}
\usepackage{enumerate}
\usepackage[hang,flushmargin]{footmisc}
\usepackage[titletoc,toc]{appendix}
\usepackage{lipsum}
\usepackage{subcaption}
\usepackage{hyperref}
\usepackage{rotating}
\usepackage{multirow}
\usepackage{multicol}
\usepackage{xfrac}
\usepackage{mathtools}
\usepackage{nccmath}
\usepackage{cite}
\usepackage{placeins}
\usepackage{tikz}
\usetikzlibrary{positioning}
\usepackage{textalpha}
\graphicspath{{figures/}} 
\usepackage[font=small,labelfont=bf]{caption}
\usepackage{wrapfig}
\usepackage{authblk}
\usepackage{cite}
\usepackage[utf8]{inputenc}
\usepackage[T1]{fontenc}
\usepackage{enumitem}
\usepackage{blindtext}

\title{ 
  \bf Late time acceleration due to generic  modification
of gravity and 
Hubble tension }
\author{Shahnawaz A. Adil\thanks{shazadil14@gmail.com}}
\affil{\small \it Department of Physics, Jamia Millia Islamia, Delhi-110025, India}
\author{ Mayukh R. Gangopadhyay\thanks{mayukh@ctp-jamia.res.in}}
\affil{\small \it Centre for Theoretical Physics, Jamia Millia Islamia, Delhi-110025, India}
	
\author[3,4,5]{M.~Sami\thanks{msami@jmi.ac.in}}

\affil[3]{\small \it Centre for Cosmology and Science Popularization(CCSP), SGT University, Gurugram 12006, India. }
\affil[4]{\small \it International Center for Cosmology, Charusat University, Anand 388421, Gujarat, India}
\affil[5]{ Center for Theoretical Physics, Eurasian National University, Astana 010008, Kazakhstan}
\author[6]{Mohit K. Sharma\thanks{mr.mohit254@gmail.com}}
\affil[6]{\small \it Department of Physics \& Astrophysics, University of Delhi, Delhi-110007, India}
\def\doi{http://doi.org}

\begin{document}
\date{}
\maketitle
\begin{abstract}
 We consider a scenario of modified gravity, which is generic to late-time acceleration, namely, acceleration in the Jordan frame and no acceleration in the Einstein frame. The possibility is realized by assuming an interaction between dark matter and the baryonic component in the Einstein frame which is removed by going to the Jordan frame using a disformal transformation giving rise to an exotic effective fluid responsible for causing phantom crossing at late times. In this scenario, past evolution is not distinguished from $\Lambda$CDM but late time dynamics is generically different due to the presence of phantom crossing that causes a monotonous increase in the expansion rate giving rise to distinctive late-time cosmic feature. The latter can play a crucial role in addressing the tension between the observed value of Hubble parameter by CMB (Cosmic Microwave Background) measurements and the local observations. We demonstrate that the Hubble tension significantly reduces in the scenario under consideration for the chosen scale factor parametrizations. The estimated age of the universe in the model is well within the observational bounds in the low and high red-shift regimes.
\end{abstract}

\section{Introduction}
Consistency of standard model of Universe necessarily asks that the hot big bang model be complemented  by two phases of accelerated expansion, namely, inflation and late time acceleration \cite{DEbook}. Observation reveals that the age of certain well known objects in the Universe exceeds the age of Universe estimated in the model assuming presence of standard matter in the Universe \cite{age,LopezCorredoira2017,age2,Damjanov,age3}. 
Since most of the contribution to the age of Universe comes from late time evolution, thereby, invoking the late time cosmic acceleration slows down the rate of Hubble expansion such that it takes more time to reach a given observed value of  the Hubble parameter, in particular, $H_0$\cite{LopezCorredoira2018}. And the latter provides with a known resolution of the puzzle. A possibility to ease such a $H_0$ tension in case of scalar tensor theories was first proposed in \cite{ballardini3, ballardini4}. Later, for non-minimally coupled scalar-tensor theories in Horndeski gravity, the tension is studied in light of the CMB and BAO datasets \cite{ballardini1}.
It is interesting to note that the late time cosmic acceleration as consistency requirement of hot big bang is supported  by direct as well as by the indirect observations \cite{Riess}-\cite{TFR}, though such a confirmation is yet to be awaited for inflation.

Broadly, there are two ways to achieve late time cosmic acceleration, namely, by adding a source term with large negative pressure (dubbed dark energy) to the energy momentum tensor in the Einstein equations \cite{Copeland,Sami,Sahni,Frieman,Caldwell,Trodden,Sami2,Peri,Frieman2,Sami-notes,Carroll,Pady,Ratra,Wetterich,Ratra2,Steinhardt} or by modifying geometry of space time $\hat{\rm a}$  {\it  la} {\it modified gravity}. As for dark energy,  a plethora of viable candidates, including quintessence, phantom fields, rolling tachyons and others have been investigated in the literature, see Ref.\cite{Copeland} for details.  The simplest model of dark energy based upon cosmological constant dubbed $\Lambda$CDM (where $\Lambda$ is the cosmological constant and CDM is the cold dark matter) is under scrutiny at present and there are strong reasons to look beyond\cite{hotens}.
The modified theories of gravity, can generally  be though as Einstein theory plus extra degrees of freedom; $f(R)$ gravity, massive gravity and Horndeski provide examples of alternative theories of large scale modification of gravity extensively discussed in the literature\cite{ijmpds}. Thus, according to the standard paradigm, late time acceleration is either sourced by the presence of an exotic matter or large scale modification of gravity. 

A third possibility that does not invoke extra degrees of freedom or exotic matter, was proposed in\cite{KHOURY2016}, see also Ref.\cite{shibesh} on the related theme. In this framework, interaction between Dark Matter(DM) and Baryonic Matter (BM) is assumed in the Einstein frame, thereby, no acceleration in this frame as the total matter still behaves as non-negative pressure fluid\footnote{If individual matter components are assumed to be of cold dark matter type then sum of both the components behaves as zero pressure fluid.}. However, an interesting situation arises in the Jordan frame connected to Einstein frame by a disformal transformation. In the Jordan frame, interaction between DM and BM is removed allowing them to adhere to conservation separately. However, an effective term gets generated in the Jordan frame which mimics an exotic fluid  providing a possibility to account for late time acceleration. In this picture, one naturally realizes a phantom crossing at late times. Interestingly, conformal coupling is disfavoured by the stability considerations whereas maximally disformal coupling can account for cosmic acceleration in Jordan frame but deceleration in Einstein frame\cite{KHOURY2016}. As a result, in this scenario, the presence of phantom phase in Jordan frame gives rise to sudden future singularity which however could be delayed to distant future by suitably parametrizing the disformal transformation\cite{shibesh}.
Let us emphasize that the said framework gives rise to late time acceleration generic to large scale gravity modification: {\it Acceleration in Jordan frame and no acceleration in Einstein frame} \cite{jwang,kBamba}. It may be noted that $f(R)$ theories fail to satisfy this criteria if local screening is properly implemented \cite{LPC1, LPC2,chamel0, chamel1, chamel2,EXTENDED,CAPOZZI}\footnote{Using a conformal transformation, $f(R)$ gravity can be transformed to Einstein gravity plus a scalar degree of freedom non-minimally coupled to matter. Since Einstein theory has been tested to great accuracy in solar system, the extra degree of freedom should locally be screened, it should only show up at large scales to account for late time acceleration. Unfortunately, proper local screening leaves no scope for late time acceleration in this framework.}. 
 
 As mentioned before, the $\Lambda$CDM is faced with a puzzle, namely, the disagreement 
 of the value of $H_0$ estimated in it using 
   the CMB likelihood  with the local  measurements. There are several proposals in the  literature where the Hubble tension is addressed from different perspectives using early time or late time cosmology \cite{Hub2}-\cite{BasilakosH0}. The latter yields higher values of the Hubble parameter compared to the ones predicted by Planck measurement which assumes  $\Lambda$CDM to be the background model. The scenario under consideration conforms to $\Lambda$CDM in the past and causes   phantom crossing around the present epoch which gives rise to  monotonously increasing Hubble parameter around the present epoch. It is interesting
    to note  that the recently carried out model independent investigations, using combined cosmological data, are in agreement with late time
     phantom crossing. Let us stress that
   this behavior is a distinguished feature of 
   the model based upon the aforesaid disformal coupling
which might naturally  provide us with a possibility to address the Hubble tension.

The plan of the paper is as follows. In section (\ref{disf}), first we describe the basics of cosmological dynamics of  disformal coupling between the DM and BM followed by discussion on the parametrizations of the scale factor. In section (\ref{method}), we describe the technical details of the analyses done using the Monte Carlo Markov Chains (MCMC) simulations using observational data. There we have shown the evidence for the proposed model of ours with respect to the standard $\Lambda$CDM cosmology using the Bayesian Information Criterion (BIC) technique. Finally a summary and the conclusions of this work are given in section (\ref{conclusion}).

\section{Disformal coupling between known components of matter} 
\label{disf}
In this section, we briefly describe the framework which includes mechanism of interaction between BM and DM in the Einstein frame which is then transformed to the Jordan frame using a disformal transformation.  Let us consider the following action in the Einstein
frame\cite{KHOURY2016}, 
\begin{equation}
\label{lag1}
\mathcal{S}=\int d^4 x \left(\frac{1}{16\pi G}\sqrt{-g}\mathcal{R}+\mathcal{L}%
_{DM}\{g_{\mu \nu }\}+\mathcal{L}_{BM}\{\widetilde{g}_{\mu \nu }\}\right),
\end{equation}
where $\widetilde{g}_{\mu \nu }$
and  $g_{\mu \nu }$ designate the Jordan and Einstein frame metric, respectively. Also, the energy momentum tensors in the Jordan and Einstein frames are defined through variational derivatives of matter actions with respect to
  $\widetilde{g}_{\mu \nu }$ and ${g}_{\mu \nu }$, respectively. In what follows, quantities with a
overhead tilde would be associated with Jordan frame.
The construct in Eq. (\ref{lag1}) implies coupling between DM and BM in the Einstein frame; obviously, their energy densities do not conserve separately but their sum does and exhibits behavior of standard matter. Consequently, in the framework under consideration, evolution would have decelerating character in the Einstein frame. The Jordan frame metric,
$\widetilde{g}_{\mu \nu }$, would be constructed from 
$g_{\mu \nu }$
 and parameters
that characterize the dark matter; we shall use disformal transformation between Einstein and Jordan frames.

For simplicity, we  assume
dark matter to be a perfect fluid which then can be represented by  a single scalar field, 
\begin{equation}
\mathcal{L}_{DM}=\sqrt{-g}P(X) ;~~ \quad X \equiv -g^{\mu \nu }\partial _{\mu }\Theta \partial _{\nu }\Theta   \, ,
\label{Ldm}
\end{equation}
where 
$\Theta$ denotes the  dark matter field. As stated above, the energy momentum tensor $T_{\mu\nu}$ of DM can be obtained by varying the Einstein frame DM action $\mathcal{S}_{DM}$ with respect to $g_{\mu\nu}$ as
\begin{equation}
T_{\mu \nu } \equiv \frac{2}{\sqrt{-g}}\frac{\delta \mathcal{S}_{DM}}{\delta g^{\mu\nu}} = 2P,_{X}\partial _{\mu }\Theta \partial _{\nu }\Theta +Pg_{\mu\nu } \, . 
\label{2a}
\end{equation}
The above expression can be written in the standard perfect fluid form, i.e.,
\begin{equation}
T_{\mu \nu
}=(\rho _{DM}+P_{DM})u_{\mu }u_{\nu }+P_{DM}g_{\mu \nu },
\end{equation}
if we identify $\rho _{DM}$, $P_{DM}$ and $u_{\mu }$ as follows:
\begin{equation}
\rho _{DM}=2P,_{X}(X)X-P(X) \, , \quad P_{DM}=P(X) \, , \quad u_{\mu }=-%
\frac{1}{\sqrt{X}}\partial _{\mu }\Theta \, . 
\label{3}
\end{equation}
The BM energy momentum tensor, on the other hand, is defined as
\begin{equation}
\widetilde{T}_{BM}^{\mu \nu }=\frac{2}{\sqrt{-\widetilde{g}}}\frac{\delta \mathcal{S}%
_{BM}}{\delta \widetilde{g}_{\mu \nu }} \, .  
\label{6}
\end{equation}
Note that in the Jordan frame, the matter components are not coupled to each other, though the dynamics
might look complicated, the energy momentum tensors of both the components
are separately conserved. In particular, 
\begin{equation}
\widetilde{\nabla}_{\mu }\widetilde{T}_{BM}^{\mu \nu }=0\text{.}  \label{6a}
\end{equation}
The coupling between DM and BM is accomplished through
Jordan frame metric $\widetilde{g}_{\mu \nu }$(appearing in $\mathcal{L}_{BM}$, Eq. (\ref{lag1})) which can be constructed from the
Einstein frame metric and  dark matter field. The Jordan and Einstein frame metrics are often thought  to be related with the each other through conformal transformation between two frames. However, such a transformation is excluded by the stability considerations. A general relation between $\widetilde{g}_{\mu \nu}$, $g_{\mu \nu }$ and $\Theta$ could be of  disformal type,
\begin{eqnarray}
\widetilde{g}_{\mu \nu} =R^{2}(X)g_{\mu \nu }+S(X)\partial _{\mu }\Theta
\partial _{\nu }\Theta ;~~S(X)\equiv \frac{R^{2}(X)-Q^{2}(X)}{X}\text{,}
\label{4A}
\end{eqnarray}%
with $R$ and $Q$ being the  arbitrary functions of X. 
Varying the action with
respect to $g_{\mu \nu }$, we obtain, Einstein equations,
\begin{eqnarray}
&& G_{\mu \nu }= 8\pi GT^{eff}_{\mu\nu} \, \\
&&T^{eff}_{\mu\nu}= T^{DM}_{\mu \nu }+QR^{3}\widetilde{T}_{BM}^{k\lambda }\left(
R^{2}g_{k\mu }g_{\lambda \nu }+(2RR,_{X}g_{k\lambda }+S,_{X}\partial
_{k}\Theta \partial _{\lambda }\Theta )\partial _{\mu }\Theta \partial _{\nu
}\Theta \right)  , 
\label{7}
\end{eqnarray}%
where the energy momentum tensor for dark matter, $T^{DM}_{\mu \nu }$  is given in (\ref{2a}). 
The second term in the expression of effective energy momentum tensor, includes coupling of dark matter and baryonic matter. The effective energy momentum tensor reduces to the sum of energy momentum tensors of the two matter components in absence of coupling, i.e., $Q=R=1$. Let us note that $T^{eff}_{\mu\nu}$ is conserved though the energy momentum tensors of individual matter components do not, thus no exotic behavior is expected in the Einstein frame.
\subsection{The FRW Evolution equations}
\label{sect:2}
Let us now specialize to spatially  homogeneous and isotropic background,
\begin{equation}
ds^{2}=-dt^{2}+a^{2}(t)\left( dx^{2}+dy^{2}+dz^{2}\right)  \label{1}
\end{equation}%
In this case, the coupling functions Q(X) and R(X) depend upon scale factor such that the Jordan frame metric takes following  form ,
\begin{equation}
\widetilde{g}_{\mu \nu }=diag\left(
-Q^{2}(a),R^{2}(a)a^{2},R^{2}(a)a^{2},R^{2}(a)a^{2}\right) \text{.}
\label{9}
\end{equation}%
Einstein equations (\ref{7}) then give rise to following evolution equations in the FRW Universe, 
\begin{equation}
3H^{2}=8\pi G \rho_{Tot}\equiv 
8\pi G\left( \rho^{eq} _{DM}\sqrt{\frac{X}{X^{eq}}}\left( \frac{%
a^{eq}}{a}\right) ^{3}-P+QR^{3}\widetilde{\rho}_{b}\right) \text{,}  \label{2}
\end{equation}%
and
\begin{equation}
2\frac{\ddot{a}}{a}+H^{2}=-8\pi G(P+P_{b})\text{,}  \label{4}
\end{equation}%
where the superscript 'eq' represents the quantity at the matter-radiation equality epoch and $P$($P_{b}$) denotes the
pressure of DM(BM) in the Einstein frame, such that $%
P_{b}\equiv QR^{3}\widetilde{P}_{b}$. 
 We assume matter to be pressure-less, namely, $\widetilde{P}_{b}\simeq 0%
\text{ and }P\ll 2XP,_{X}\label{21}$. 
Since baryonic matters adheres to conservation in the Jordan frame, we have,
\begin{equation}
\tilde{\rho}_b=\frac{\rho_b^{eq}}{R^3}\left(\frac{ a^{eq}}{a} \right)^3  \, .
\end{equation}
Under the said assumptions on $\tilde{P}_b$ and $P$, the RHS of (\ref{4}) vanishes which implies that background in the Einstein frame is matter dominated, i.e., $a(t)\sim t^
{2/3}$ throughout. The total matter density in the Einstein frame then acquires the form,
\begin{equation}
\rho_{Tot}(a)=\left(\rho^{eq}_{DM} \sqrt{\frac{X}{X_{eq}}}+Q(X)\rho^{eq}_{b}\right)\left(\frac{ a_{eq}}{a}\right)^3 \, ,
\end{equation}
where  the quantity within the brackets should be constant for an arbitrary function $Q(X)$ as the background in the Einstein frame is matter dominated as it should be.

Let us note that coupling is conformal if $Q=R$ otherwise it is disformal. It turns out that system is plagued with with instability in case of conformal coupling\cite{KHOURY2016}.
In what follows, without the loss of generality, we shall adhere to the maximally disformal case, namely, $Q(a)\equiv 1$ leaving with a single  function $R$ to deal with. The function, $R(a)$,  should be such that the thermal history, known to great accuracy,  be left intact. Thereby, the physical scale factor $\widetilde{a}=R(a) a$
should agree with $a$ throughout the entire history and should 
 grow
sufficiently fast only at late stages such that the physical scale factor $\widetilde{a}$ in Jordan
frame experiences acceleration\footnote{Recall that we desire to have acceleration in the Jordan frame,
($\ddot {\widetilde{a}}>0$) and deceleration in the Einstein frame($\ddot {{a}}<0$ ). If
 $R$ is concave up, the growth of $R$ at late times might compensate the effect of deceleration in $a(t)$
making $\ddot{\widetilde{a}}$ positive in the expression, $ \ddot{\widetilde{a}}=\ddot{R}a+2\dot{R}\dot{a}+R\ddot{a}$.
 }. In what follows, we shall use convenient the parametrizations for the scale factor, keeping in mind, the mentioned phenomenological features.
\subsection{Scale factor: Polynomial parametrization} 
 
In order to proceed further, we need to specify the function $R(a)$ that  conforms to the said requirements. To this effect, we shall use the following parametrizations, namely, the polynomial and exponential ones, for a concrete realization,
\begin{eqnarray}
\label{m1}
&&  a(\widetilde{a})=\widetilde{a}+\alpha \widetilde{a}^{2}+\beta \widetilde{a}^{3}~~~ \text{(Polynomial)}, \\
&&  a(\widetilde{a})=\widetilde{a}e^{\alpha \widetilde{a}}~~~~~~~~~~~~~~~~\text{(Exponential)}
\label{m2}
\end{eqnarray}
where $\alpha$ and $\beta$ are real parameters.

In the analysis to follow, we would require to present  the cosmological parameters in terms of redshifts in
both frames\cite{shibesh}\footnote{Let us note that the the physical scale factor can be normalized to one,  $\widetilde{a}_{0}=1$, however, $a_{0}=1+\alpha+\beta\neq 1$ (Polynomial parametrization  (\ref{m1})) and $a_{0}=e^\alpha \neq 1$ (Exponential parametrization, (\ref{m2})).}, 
\begin{equation}
\widetilde{a}=\frac{\widetilde{a}_{0}}{1+\widetilde{z}}\text{ , \ \ }a=\frac{a_{0}}{1+z}%
\text{,}  \label{5}
\end{equation}%

Using  (\ref{5}), the Hubble parameter, in case of model (\ref{m1}) can be cast in terms of redshift $\widetilde{z}$ as ,
\begin{equation}
\widetilde{H}(\widetilde{z})=\widetilde{H}_{0} F_{(pol)}(\alpha,\beta,\tilde{z})\equiv \widetilde{H}_{0}\frac{(1+\alpha +\beta )^{\frac{1}{2}%
}(1+2\alpha +3\beta )\left( 1+\widetilde{z}\right) ^{\frac{9}{2}}}{\left[ \left(
1+\widetilde{z}\right) ^{2}+\alpha \left( 1+\widetilde{z}\right) +\beta \right] ^{%
\frac{1}{2}}\left[ \left( 1+\widetilde{z}\right) ^{2}+2\alpha \left( 1+\widetilde{z}%
\right) +3\beta \right] }\text{,}  \label{a3}
\end{equation}%
which can be written through effective fractional density parameters (see Appendix).
The corresponding effective (total) equation of state parameter is given by,
\begin{equation}
\widetilde{w}_{eff}(\widetilde{z}) = -\frac{2\dot{\widetilde{H}}}{~3\widetilde{H}^2} = \frac{\alpha \left( 5+6\alpha +5\widetilde{z}\right)
(1+\widetilde{z})^{2}+\beta (14+23\alpha +14\widetilde{z})(1+\widetilde{z})+18\beta ^{2}%
}{3\{(1+\widetilde{z})^{2}+\alpha (1+\widetilde{z})+\beta \}\{(1+\widetilde{z}%
)^{2}+2\alpha (1+\widetilde{z})+3\beta \}}\text{.}  \label{a5}
\end{equation}
The dark energy (DE) equation of state $\widetilde{w}_{de}$ can then be obtained from the following relation:
\begin{equation} \label{wtotal}
\widetilde{w}_{eff}(\widetilde{z}) = \widetilde{w}_{M}\Omega^{(0)}_{M eff} + \widetilde{w}_{de}(\widetilde{z}) \, \Omega^{(0)}_{DE} \Rightarrow \widetilde{\omega}_{de}(z)=\widetilde{\omega}_{eff}(\tilde{z})/\Omega^{(0)}_{DE} \, .
\end{equation}
where $\Omega^{(0)}_{M eff}$ and $\Omega^{(0)}_{DE}$ are the effective matter and DE density parameters, respectively, whereas, $ \widetilde{w}_{M}$($\widetilde{w}_{M}=0$) is the equation of state parameter for matter. 
It is also important to mention that for $\alpha=-0.1523$ and $\beta=-0.0407$, the equation of state parameter for this parametrization approaches  its $\Lambda$CDM limit i.e., $w_{de}\to -1$ at the present epoch. 

\subsection{Scale factor : Exponential parametrization}

The Friedmann equation and the effective equation of state parameter for parametrization (\ref{m2}), can be written as
\begin{equation} \label{b3}
\widetilde{H}(\widetilde{z})= \frac{\widetilde{H}_0 (1+\alpha)(1+\widetilde{z})^{5/2}}{1+\widetilde{z}+\alpha}\exp{\left(\frac{3\widetilde{z}\alpha}{2(1+\widetilde{z})}\right)}\equiv H_0 F_{(exp)}(\tilde{z},\alpha) \, ,
\end{equation}
\begin{equation}
\widetilde{w}_{eff}(\widetilde{z})=\frac{5\alpha (1+\widetilde{z})+3\alpha ^{2}}{3(1+%
\widetilde{z})\left[ (1+\widetilde{z})+\alpha \right] }\text{ .}  \label{b5}
\end{equation}
Let us note that in both cases, the dark energy equation of state parameter 
might assume super negative values at late times with a generic behavior embodied with phantom crossing. It is worth noting that such a phenomenon can not be realized by quintessence field; one needs at least two scalar fields to mimic the phantom crossing. It is interesting that the presence of  disformal coupling between known components of matter inevitably gives rise to  the mentioned behavior. It should also be emphasized that the coupling between DM and BM is removed in the Jordan frame but the Einstein-Hilbert action gets modified. However, the modification of gravity under consideration is not accompanied by extra degree(s) of freedom but allows to realize super  acceleration at late times. Last but not least, acceleration in this framework is generically caused by modification of gravity 
$\hat{\rm a}$  {\it  la} {\it acceleration in Jordan frame and deceleration in Einstein frame.}

\section{Methodology and Analysis}
\label{method}
In this section, we shall perform the parametric estimation for both parametrizations (\ref{m1}) and (\ref{m2}) using the late time observational data. In order to do so, and to predict the physical significance for both the parametrizations, one needs to consider the standard form of the Friedmann equations. Since, the parametrization (\ref{m1}) is already formulated in the standard form (see appendix (\ref{app-p1-dm})), it is also necessary and desirable to obtain the similar form for the parametrization  (\ref{m2}). However, in this case, due to the exponential function, it is not tenable and therefore, we resort to the following anzats  for the Friedmann equation,
\begin{eqnarray} \label{p1-standard}
\left(\frac{\widetilde{H}(\widetilde{z})} {\widetilde{H} _0}\right)^2 &=& (1+\alpha)^A (1+\widetilde{z})^3 + B\, e^{C \, \widetilde{z}} ,\\
\widetilde{H}(\widetilde{z}) & = & \widetilde{H} _0 F_{(exp)}(\tilde{z},\alpha),
\end{eqnarray}
where $A$, $B$ and $C$ are constants. Note that here we can identify $(1+\alpha)^A$ as the effective matter density ($\Omega^{(0)}_{M eff}=(1+\alpha)^A$) and $B$ to be the effective DE density ($B=\Omega^{(0)}_{DE}$) at the present epoch. Also, the condition $\Omega^{(0)}_{M eff} + \Omega^{(0)}_{DE}=1$, implies that $(1+\alpha)^A+B=1$. By expressing $B$ in terms of $A$, we are finally left with only $A$ and $C$ to fit $F_{exp}(\tilde{z},\alpha)$ with (\ref{p1-standard}). 
Let us note that the fitting is such that
at the present epoch one gets back $\widetilde{H}(\widetilde{z}) = \widetilde{H_0}$, as usual. Whereas, at any $\widetilde{z} \neq 0$, one always sees the presence of $\alpha$ in the Friedmann equation. Using numerical techniques (in particular, non linear model fitting in the Mathematica software), we obtain $A= 3.4185$ and $C = 0.2896$.

\begin{figure}[!ht]
\centering
\includegraphics[height=2.2in,width=3.2in]{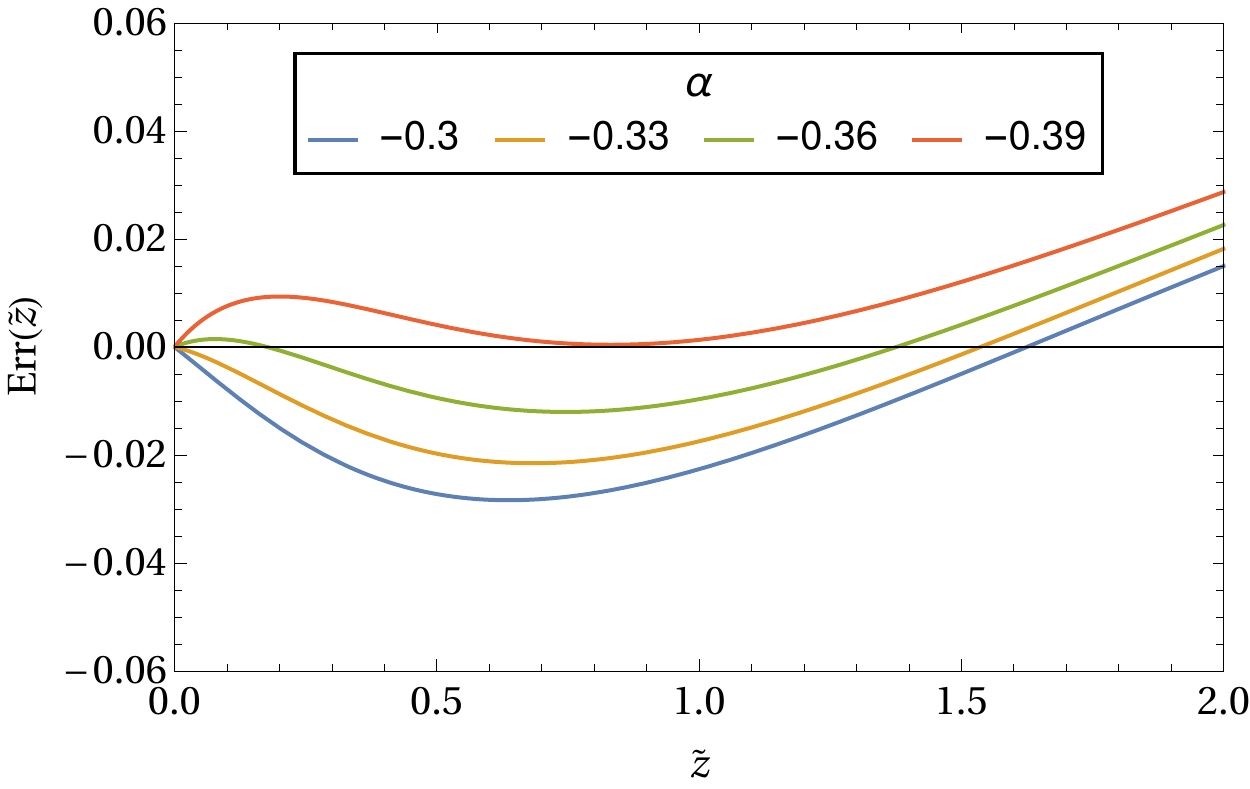}
 \caption{The figure shows the relative error between fitted and the exact Friedmann equation with $z\in[0,2]$ for different values of $\alpha$. }
\label{err-fig}
\end{figure}

In fig. (\ref{err-fig}), we plot the relative error between the fitted parametrization (\ref{p1-standard}) with respect to the Eq. (\ref{b3}). In particular, we plot the error function $Err \equiv (E_{fit} - E)/E$ (where $E \equiv \widetilde{H}/\widetilde{H_0}$). In this plot we can see that the fitting is reasonably good atleast up to redshift ($\widetilde{z} \leq 2$) for different possible values of $\alpha$.

After formulating the Friedmann equation (\ref{b3}) in the standard form (\ref{p1-standard}), we can extract out the effective DE equation of state $\widetilde{w}_{de}(\widetilde{z})$ from Eq. (\ref{wtotal}) and can constrain it by using the observational data. Since, we have already expressed the effective equation of state of the system (see Eq. (\ref{b5})), then it follows that
\begin{equation}
\widetilde{w}_{de}(\widetilde{z}) = -\frac{\alpha  \, e^{-0.2896 \widetilde{z}} \left[\alpha +1.6667(\widetilde{z}+1)\right]}{\left[(\alpha +1)^{3.4185}- 1\right] (\widetilde{z}+1) (\alpha +\widetilde{z}+1)} \, ,
\end{equation}
where we have used the fitted value of $B$ and $C$ for $\Omega^{(0)}_{DE}$. Note that for $\alpha = -0.3896$, the exponential parametrization approaches its $\Lambda$CDM model i.e., $\widetilde{w}_{de}(0)=-1$ at the present epoch.

\subsection{Data}

We have combined three set of data for our analysis.  We have taken into account the distance modulus measurement of type Ia supernovae (SNIa), observational  Hubble data (OHD) and angular diameter distances measured using water megamasers. The analysis is done in the following way:

\subsubsection{Supernova Type Ia - Pantheon Sample}

\begin{table}[!ht] 
\centering
\begin{tabular}{|ccccccc|}
\hline
\multicolumn{1}{|c}{$z$} & \multicolumn{1}{c}{$E(z)$} & \multicolumn{5}{c|}{Correlation matrix} \\
\hline
 $0.07$& $0.997\pm0.023$ & $1.00$  &   &   &   &     \\
 $0.20$& $1.111\pm0.020$&  $0.39$ &  $1.00$ &   &   &    \\
 $0.35$& $1.128\pm0.037$ & $0.53$  & $-0.14$  & $1.00$  &   &    \\
 $0.55$& $1.364\pm0.063$ &  $0.37$ &  $0.37$ & $-0.16$  & $1.00$  &   \\
 $0.90$& $1.52\pm0.12$&  $0.01$ &  $-0.08$ &  $0.17$ &  $-0.39$ & $1.00$   \\
 \hline
\end{tabular}
\caption{$z$ versus $E(z)$ data together with the correlation matrix between data points obtained from the Supernovae Type-1a Pantheon data \cite{amendola}}.
\label{tabsn}
\end{table}
 
In \cite{riess}, Reiss et al. had presented Hubble rate $E(z_i) = H(z_i)/H_0$ 
\footnote{Since, we use ($\widetilde{}$) to denote our Jordan-frame cosmological parameters, therefore, in our analysis the observed redshift is denoted as $\widetilde{z}$ and the observed Hubble rate can be expressed as $E(\widetilde{z}_i) = \widetilde{H}(\widetilde{z}_i)/\widetilde{H}_0$ } 
data points for six different redshifts in the range $z \in [0.07,1.5]$ effectively compressing the information of the - Pantheon compilation \cite{pantheon} and $15$ Sn1a at $z >1$ of the CANDELS and CLASH Multi-Cycle Treasury (MCT) programs obtained by the Hubble Space Telescope (HST), 9 of which are at $1.5 <z <2.3$. We have used this data and is enlisted in Table (\ref{tabsn}). Following the arguments given in  \cite{amendola} we have also omitted the data point at $z=1.5$ from the Table 6 in \cite{riess}. 

Theoretically, the dimensionless Hubble rate $\tilde{h}$ is defined as $\frac{\widetilde{H}(\widetilde{z})}{\widetilde{H_0}}$ and hence $\chi^2$ for the Supernova data is calculated as
\begin{equation}
\chi^2_{\rm{SN}}=\sum_{i,j}\left(E_i-\tilde{h}_i\right)\cdot c_{ij}^{-1}\cdot \left(E_i-\tilde{h}_i\right) \, ,
\end{equation}
where $c_{ij}$ is the correlation matrix given in table (\ref{tabsn}).

\subsubsection{Observational Hubble Data(OHD)}

We use the observational measurements of Hubble parameter at different redshifts in the range $0.07< z <1.965$. In particular, we consider a compilation of $31$ $H(z)$ measurements obtained from the cosmic chronometric method \cite{Morescoetal2012} and the compiled data set is presented in Table (\ref{H_z}).
\begin{table}[!htp]
\renewcommand{\arraystretch}{1}
\centering
\begin{minipage}{0.4\textwidth}
\begin{tabular}{|c|c|c|c|} 
  \hline
$z$ &   $H_{obs}$     &  $\sigma_{H}$ & Reference \\ \hline
0.07	&	69	&	19.6	& 	\cite{Zhangetal2014}\\
0.09	&	69	&	12	& 	\cite{Simonetal2005} \\
0.12	&	68.6 &	26.2	& 	\cite{Zhangetal2014} \\
0.17	&	83	&	8	& 	\cite{Simonetal2005}\\
0.179	&	75	&	4	& 	\cite{Morescoetal2012} \\
0.199	&	75	&	5	& 	\cite{Morescoetal2012}\\
0.2	    &	72.9 &	29.6	&  \cite{Zhangetal2014} \\
0.27	&	77	&	14	&   \cite{Simonetal2005} \\
0.28	&	88.8 &	36.6	&   \cite{Zhangetal2014} \\
0.352	&	83	&	14	&  \cite{Morescoetal2012}\\
0.38	&	81.9 &	1.9	&   \cite{Alametal2017}\\
0.3802	&	83	&	13.5	&  \cite{Morescoetal2016} \\
0.4	        &	95	&	17	&   \cite{Simonetal2005} \\
0.4004	&	77	&	10.2	&   \cite{Morescoetal2016} \\
0.4247	&	87.1	&	11.2	&  \cite{Morescoetal2016} \\
0.4497	&	92.8	&	12.9	&   \cite{Morescoetal2016}\\
0.47        &       89  &   50      &  \cite{Ratsimbazafyetal2017} \\
\bottomrule
\end{tabular}
\end{minipage}  
\begin{minipage}{0.4\textwidth}
\begin{tabular}{|c|c|c|c|}
  \hline
$z$ &   $H_{obs}$ &  $\sigma_{H}$ & Reference \\ \hline
0.4783	&	80.9	&	9	&   \cite{Morescoetal2016}\\
0.48	&	97	&	62	&  \cite{Ratsimbazafyetal2017} \\
0.51	&	90.8	&	1.9	&   \cite{Alametal2017} \\
0.593	&	104	&	13	&   \cite{Morescoetal2012} \\
0.61	&	97.8	&	2.1	&   \cite{Alametal2017} \\
0.68	&	92	&	8	&   \cite{Morescoetal2012} \\
0.781	&	105	&	12	&  \cite{Morescoetal2012} \\
0.875	&	125	&	17	&   \cite{Morescoetal2012} \\
0.88	&	90	&	40	&    \cite{Ratsimbazafyetal2017} \\
0.9	        &	117	&	23	&   \cite{Simonetal2005} \\
1.037	&	154	&	20	&  \cite{Morescoetal2012} \\
1.3	        &	168	&	17	&    \cite{Simonetal2005} \\
1.363	&	160	&	33.6	&  \cite{Moresco2015}\\
1.43	&	177	&	18	&   \cite{Simonetal2005} \\
1.53	&	140	&	14	&  \cite{Simonetal2005} \\
1.75	&	202	&	40	&   \cite{Simonetal2005} \\
1.965	&	186.5	&	50.4 &   \cite{Moresco2015} \\
\bottomrule
\end{tabular}
\end{minipage}
\caption{Observational hubble data $H_{obs}$ in the units of km s$^{-1}$ Mpc$^{-1}$ with their corresponding uncertainties $\sigma_{H}$ for various redshifts.}
\label{H_z}
\end{table}

Theoretically we calculate Hubble parameter at different redshifts for our models in different parametrizations as  $\widetilde{H}(\widetilde{z})=\widetilde{H_0}\times E(\widetilde{z})$, where $\widetilde{H_0}$ is the present value of the Hubble parameter. We rescale the present value of the Hubble parameter $\widetilde{H_0}$ to $\widetilde{h}$ where $\widetilde{h}=\widetilde{H_0}/$(100 km sec$^{-1}$ Mpc$^{-1}$) and treat $\widetilde{h}$ as a free parameter in our analysis.The $\chi^2$ for the Hubble parameter measurements is
\begin{equation}
\chi^2_H=\sum_{i}\bigg[\frac{H_i^{th}-H_i^{obs}}{\sigma^{H}_i}\bigg]^2
\end{equation}
where  $H^{\rm{obs}}$ is the observed data from the cosmic chronometer measurements and galaxy distribution measurements given in  Table~\ref{H_z}, $H^{\rm{th}}$ is the Hubble parameter of our model, $\sigma_i^{H}$ is the standard deviation at different redshift in Table~\ref{H_z}.

\subsubsection{Masers Data :}

We have also used the angular diameter distances measured by Megamaser Cosmology Project using water megamasers. The Megamaser data we have used in this analysis is given in Table 5 of \cite{Jarahetal2017}. For the sake of completeness we repeat the Table here \ref{masertab}. $\chi^2_{\rm{masers}}$ is defined conventionally as
\begin{equation}
\chi^2_{\rm{mas}}=\sum_{i}\bigg[\frac{D_{Ai}^{th}-D_{Ai}^{obs}(z_i)}{\sigma^{D}_i}\bigg]^2
\nonumber
\end{equation}
where $D_A^{th}$ is the angular diameter distance for the cosmological model calculated using definition :
\begin{equation}
D_A(z)=\frac{1}{1+z}\int_0^z\frac{dz^\prime}{H(z^\prime)}
\end{equation}
\begin{table*}[!hbt]
\centering
\caption{Megamaser data}
\begin{tabular}{|c|l|l|l|l|}
\hline
Maser&Redshift&Constraint&Ref.\\
\hline\hline
UGC 3789&$z=0.0116$&$\frac{D_A(0.0116)}{\rm{Mpc}}=49.6\pm 5.1$&\cite{Reidetal2013}\\
\hline
NGC 6264&$z=0.0340$&$\frac{D_A(0.0340)}{\rm{Mpc}}=144\pm 19$&\cite{Kuoetal2013}\\
\hline
NGC 5765b&$z=0.0277$&$\frac{D_A(0.0277)}{\rm{Mpc}}=126.3\pm 11.6$&\cite{Gaoetal2016}\\
\hline
\end{tabular}
\label{masertab}
\end{table*}

\subsection{Analysis}

We have performed the statistical analysis with Bayesian inference technique, which is extensively used for parameter estimation in cosmological models. According to this statistics the posterior probability distribution function of the model parameters is proportional to the likelihood function and the prior probability of the model parameters. To estimate the parameters, the likelihoods used are commonly  multivariate Gaussian likelihoods given by.
\begin{equation}
{\cal L} (\Theta)\propto {\rm exp}\left[-\frac{\chi^2(\Theta)}{2}\right] \, ,
\end{equation}
where $\Theta$ is a set of model parameters. In this analysis we have used uniform prior. So the posterior probability is proportional to ${\rm exp}[-\frac{\chi^2(\Theta)}{2}]$. Thus a minimum $\chi^2(\Theta)$ will ensure a maximum likelihood or a maximum posterior probability. In our analysis the Gaussian likelihood is given by
\begin{equation}
{\cal L}\propto {\rm exp}\left[-\frac{\chi^2_{\rm TOT}}{2}\right] \, ,
\end{equation}

where $\chi^2_{\rm TOT}= \chi^2_{\rm SN}+\chi^2_{\rm H}+\chi^2_{\rm masers}$.

\begin{table*}[!htb]
\centering
\begin{tabular}{|c|c|c|} 
		\hline
		parameters & \multicolumn{2}{|c|}{Priors} \\
		\hline
	      & Exponential & Polynomial \\ 
		\hline
		 $\alpha$ & [-1 , 0 ] & [-0.5 ,0 ] \\
		 \hline 
		 $\beta$ &  - & [ -0.2, 0 ] \\
		 \hline
		 $\widetilde{h}$ & [0.55 ,0.85 ] & [0.55 , 0.85] \\
		 \hline		
\end{tabular}
\caption{Priors for the MCMC parameters}
\label{priorTab}
\end{table*}
We have constrained different set of parameters for each parametrizations in our analysis. For the Polynomial parametrization the constrained parameters are $\alpha$, $\beta$ and $\widetilde{h}$. Similarly for the exponential parametrization we have constrained $\widetilde{h}$ and $\alpha$. As mentioned earlier in this analysis we have used uniform priors. We have enlisted the priors that we have used for each parametrizations in table \ref{priorTab}. We have used Markov Chain Monte Carlo (MCMC) method to make the parameter estimation. For this purpose we have used MCMC sampler, EMCEE,\cite{Goodmanetal2010,Foremanetal2013} in PYTHON. We have studied these MCMC chains using the program GetDist \cite{Lewis}.


\subsection{Constraints on Hubble parameter and dark energy equation of state parameter: Phantom crossing and Hubble tension }

\begin{figure*}[!ht]
\centering
\includegraphics[scale=0.8]{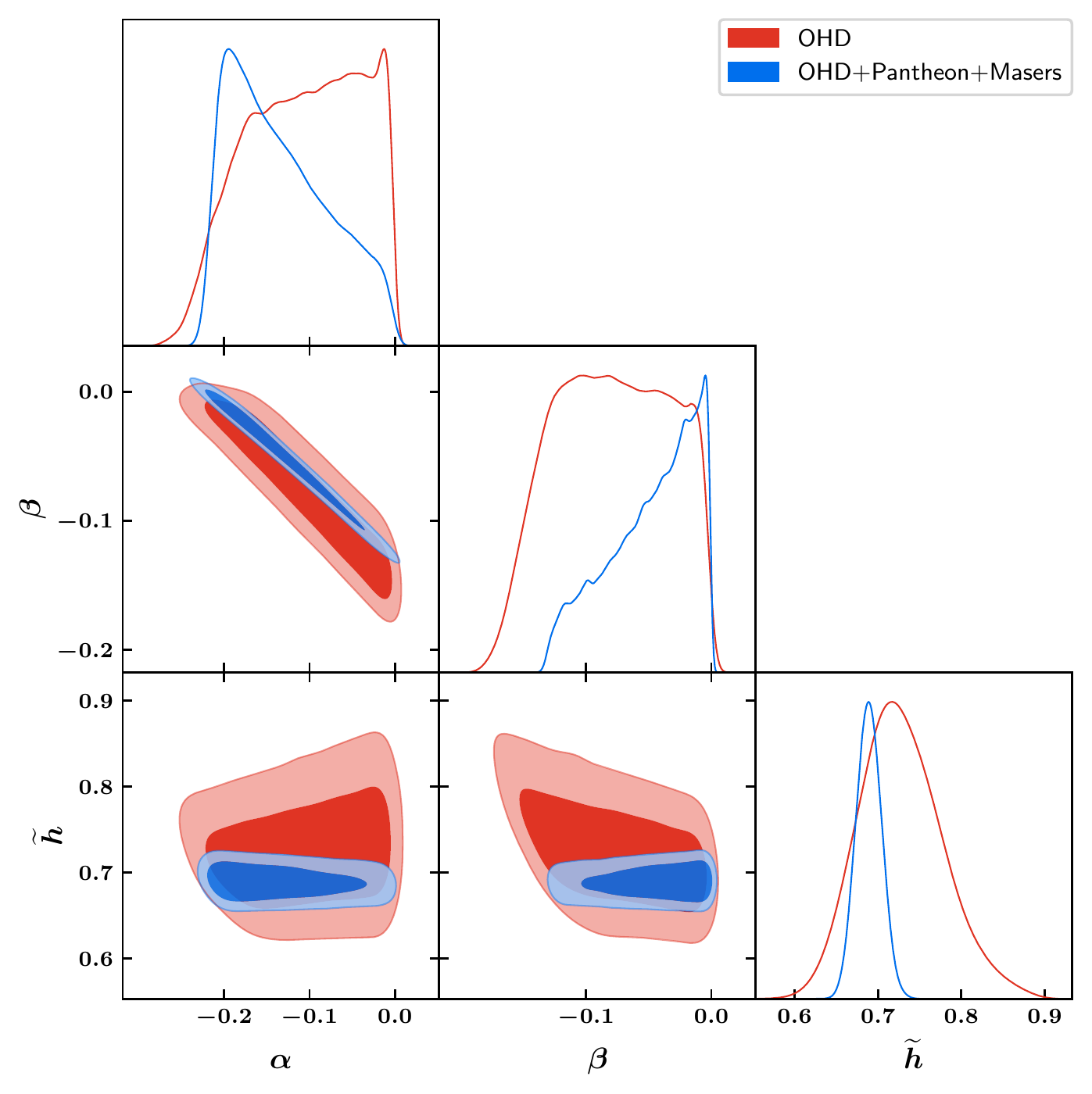}
\caption{Polynomial: $2\sigma$ contour levels between $\alpha$, $\beta$ and $\widetilde{h}$ for OHD and its combinations with Pantheon+Masers.}
\label{fig: poly_contour}
\end{figure*}

\begin{figure*}[!ht] 
\centering
\begin{subfigure}{0.495\linewidth} \centering 
   \includegraphics[height=5.4cm,width=6.4cm]{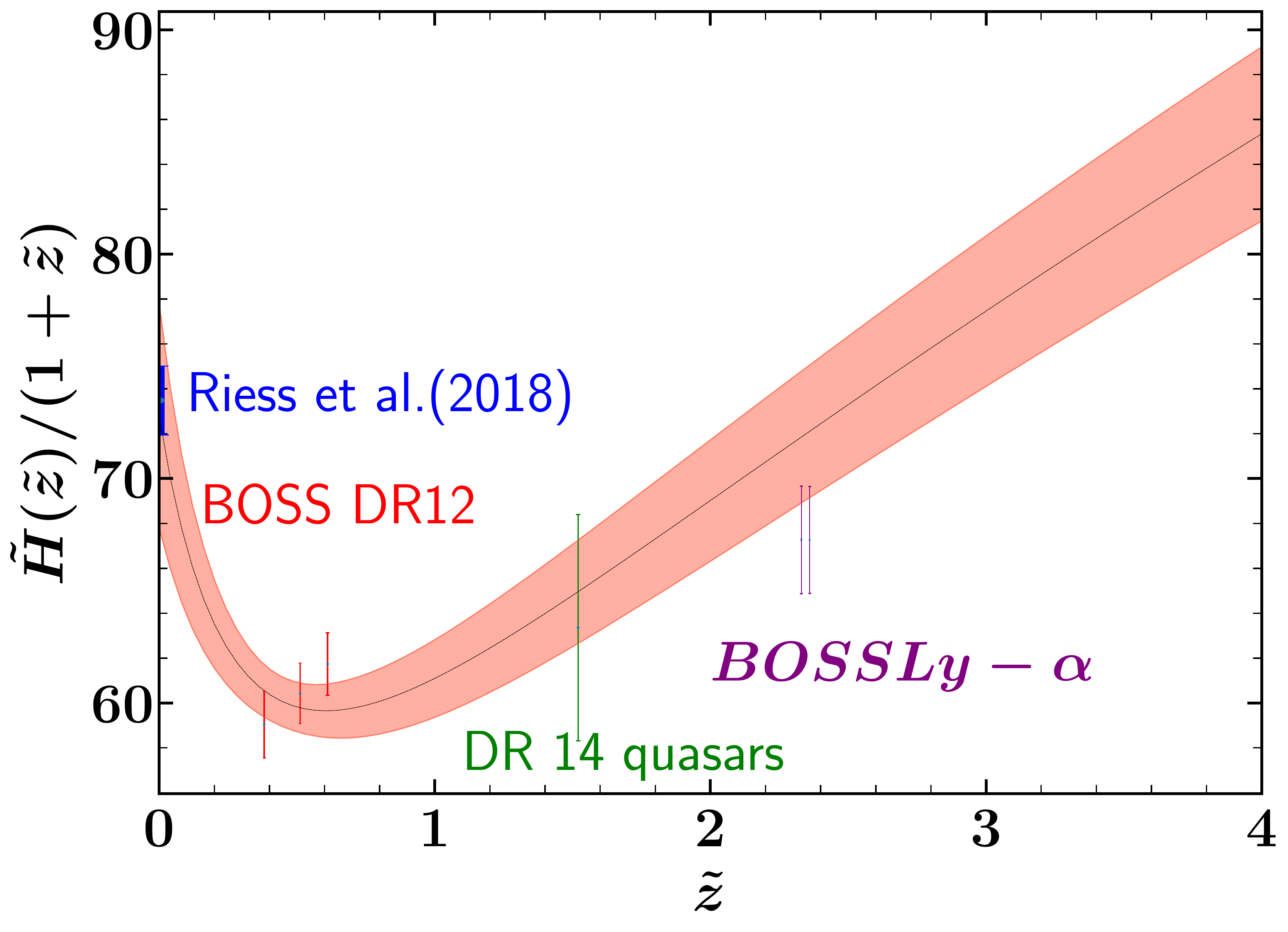}
   \caption{} \label{poly-H-a}
\end{subfigure}
\begin{subfigure}{0.495\linewidth} \centering
    \includegraphics[height=5.4cm,width=6.4cm]{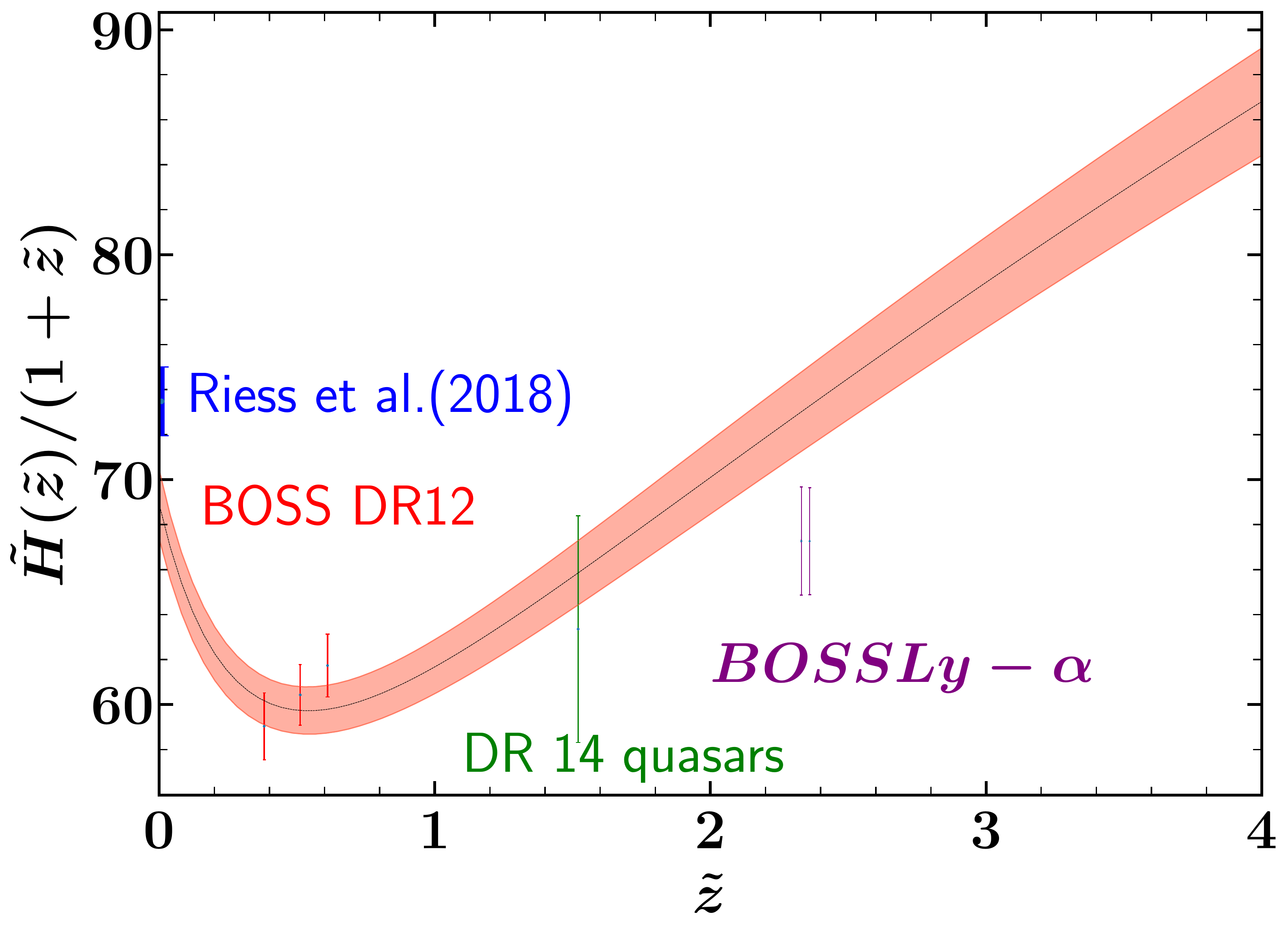}
    \caption{} \label{poly-H-b}
\end{subfigure}
\caption{Polynomial: Figures (\ref{poly-H-a}) and (\ref{poly-H-b}) depicts the evolution of $\widetilde{H}(\widetilde{z})/(1+\widetilde{z})$ with $\widetilde{z} \in [0,4]$ for the datasets OHD and OHD+Pantheon+Masers, respectively. The dark line represents the best-fit and the shaded region corresponds to the $1\sigma$ limit. }
\label{fig:H_poly}
\end{figure*}

\begin{figure*}[!ht] 
\centering
\begin{subfigure}{0.495\linewidth} \centering 
   \includegraphics[height=5.4cm,width=7cm]{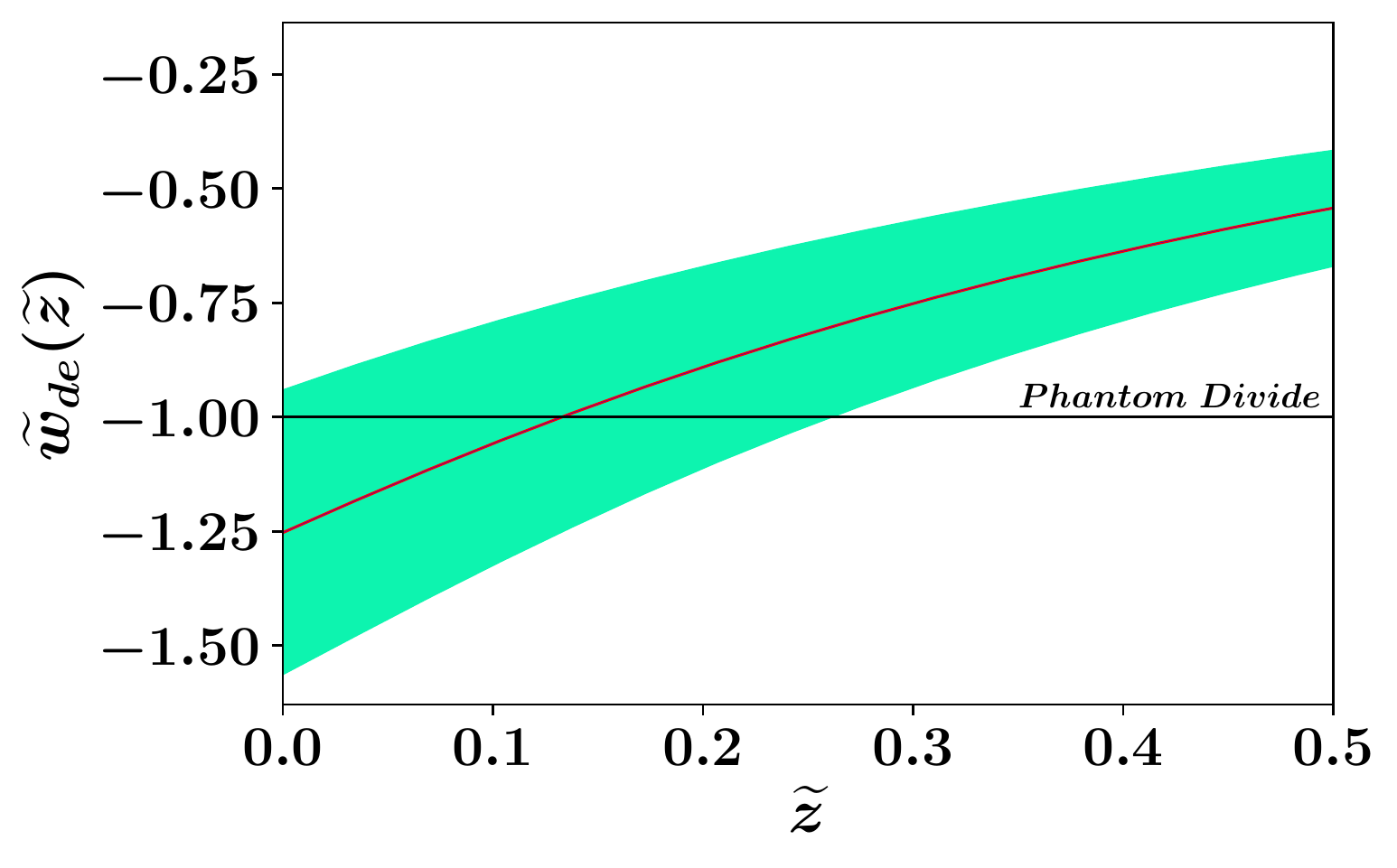}
   \caption{} \label{poly-wde-a}
\end{subfigure}
\begin{subfigure}{0.495\linewidth} \centering
    \includegraphics[height=5.4cm,width=7cm]{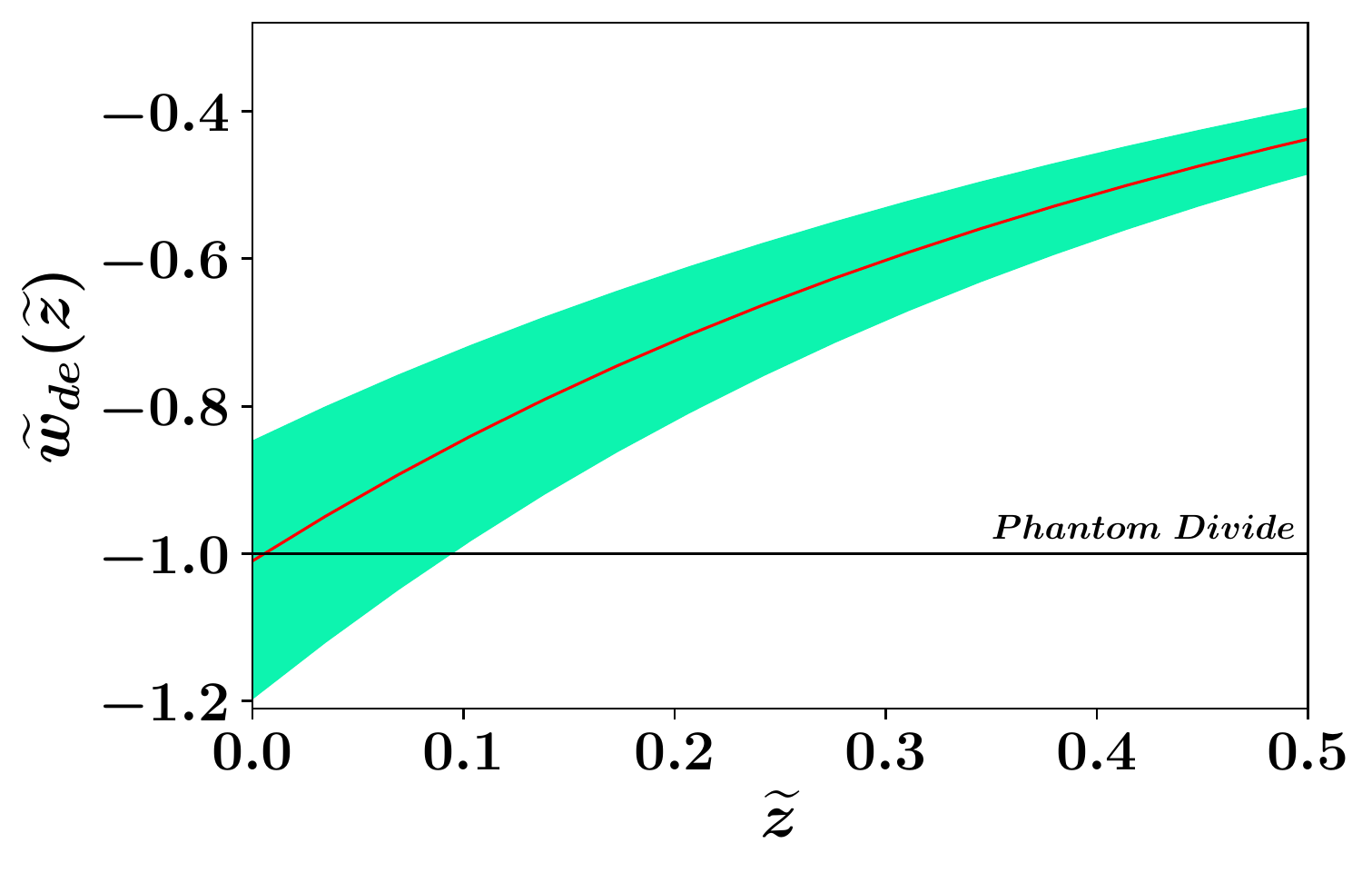}
    \caption{} \label{poly-wde-b}
\end{subfigure}
\caption{Polynomial: Figures (\ref{poly-wde-a}) and (\ref{poly-wde-b}) depicts the evolution of $\widetilde{w}_{de}(\widetilde{z})$ with $\widetilde{z} \in [0,4]$ for the datasets OHD and OHD+Pantheon+Masers, respectively. The dark line represents the best-fit and the shaded region corresponds to the $1\sigma$ limit.
}
\label{fig:wde_poly}
\end{figure*}

From the obtained values of parameters $\alpha$, $\beta$ and $\widetilde{h}$, from our MCMC simulation, upto $2\sigma$ (shown in fig.\ref{fig: poly_contour}), we can plot the evolutionary profiles of $\widetilde{H}(\widetilde{z})$ and $\widetilde{w}_{de}(\widetilde{z})$ 
\footnote{While plotting the desired functional profile we have again run the simulation where the parameters are not kept fixed to their best-fits, but instead by using their values up to $3\sigma$ level. }.

\begin{table*}[ht]
\centering
\renewcommand{\arraystretch}{1.2}
\begin{tabular}{|c|c||c|c|}  %
	\hline
	&   \multicolumn{2}{|c|}{Parametrizations} & \\
	\cline{2-3}
	Observational &  Polynomial & Exponential & $\Lambda$CDM \\
	dataset &  Best-fit($\pm 1\sigma$) &  Best-fit($\pm 1\sigma$) & \\
	\hline
	  & $\widetilde{h} =  0.7279_{-0.05}^{+0.05}$  & $\widetilde{h} = 0.671_{-0.029}^{+0.029} $  & $\widetilde{h} = 0.6770^{+0.030}_{-0.030}$  \\
 	 OHD & $ \alpha=-0.101^{+0.07}_{-0.077}$ &  $\alpha=-0.299^{+0.043}_{-0.042}$ & $\widetilde{\Omega}_M = 0.3249^{+0.064}_{-0.059}$  \\
	 & $\beta = -0.078^{+0.051}_{-0.049}$ &  - & \\
	\hline
	 & $\widetilde{h}$ =  $0.689_{-0.015}^{+0.015}$  & $\widetilde{h}$ = $0.677_{-0.007}^{+0.007} $ & $\widetilde{h}=0.6683^{+0.026}_{-0.026}$\\
OHD+Pantheon +Masers	& $\alpha= -0.145^{+0.078}_{-0.051}$   & ${\alpha}$ =  $-0.335_{-0.017}^{+0.016}$ & $\widetilde{\Omega}_M = 0.3440^{+0.061}_{-0.054}$\\
& $\beta = -0.041^{+0.029}_{-0.047}$ &  - & \\
	\hline\hline
\end{tabular}
\caption{Best-fits with their $1\sigma$ levels for polynomial and exponential parametrizations, and for the $\Lambda$CDM model from OHD and OHD+Pantheon+Masers datasets. }
\label{table1}
\end{table*}

In fig.\ref{fig: poly_contour}, we have plotted the obtained parametric dependence between $\alpha$, $\beta$ and $\widetilde{h}$ using OHD and its combination with Pantheon and Masers. From this figure, one can note that the combination of all data set significantly reduces the errors bars on $\widetilde{h}$ as compared to only OHD data set. The obtained results are given in table (\ref{table1}) where we show that only OHD data set gives $\widetilde{h}$ significantly larger than the combined data set. In particular, we have not found any significant Hubble tension for OHD, even for the combined data set, the tension reduces to $1.3\sigma$ level. The reduction of this tension can be attributed to the phantom crossing taking place in case of the polynomial parametrization.


The corresponding evolution of $\widetilde{w}_{de}(\widetilde{z})$ and $\widetilde{H}(\widetilde{z})$ upto $1\sigma$ are shown in figures (\ref{fig:H_poly}) and (\ref{fig:wde_poly}), respectively. In the fig (\ref{fig:H_poly}), one can see that due to large $1\sigma$ deviations, the only OHD data set does not show a tension with the Riess et. al. \cite{riess2018} and BOSSLy-$\alpha$ \cite{Ly1,Ly2},
whereas the combined data set by giving rise to comparatively small $\widetilde{h}$ shows a significant tension with the Riess et. al. (\ref{fig:H_poly}). On the other hand, in fig. (\ref{fig:wde_poly}), we show that both of the data sets gives rise to a consequential amount of phantom crossing at present epoch. It is interesting to note that this feature, without adding any extra degrees of freedom, is solely happening due to the coupling between two components of matter (BM and DM). 


\begin{figure*}[!ht]
\centering
\includegraphics[scale=0.86]{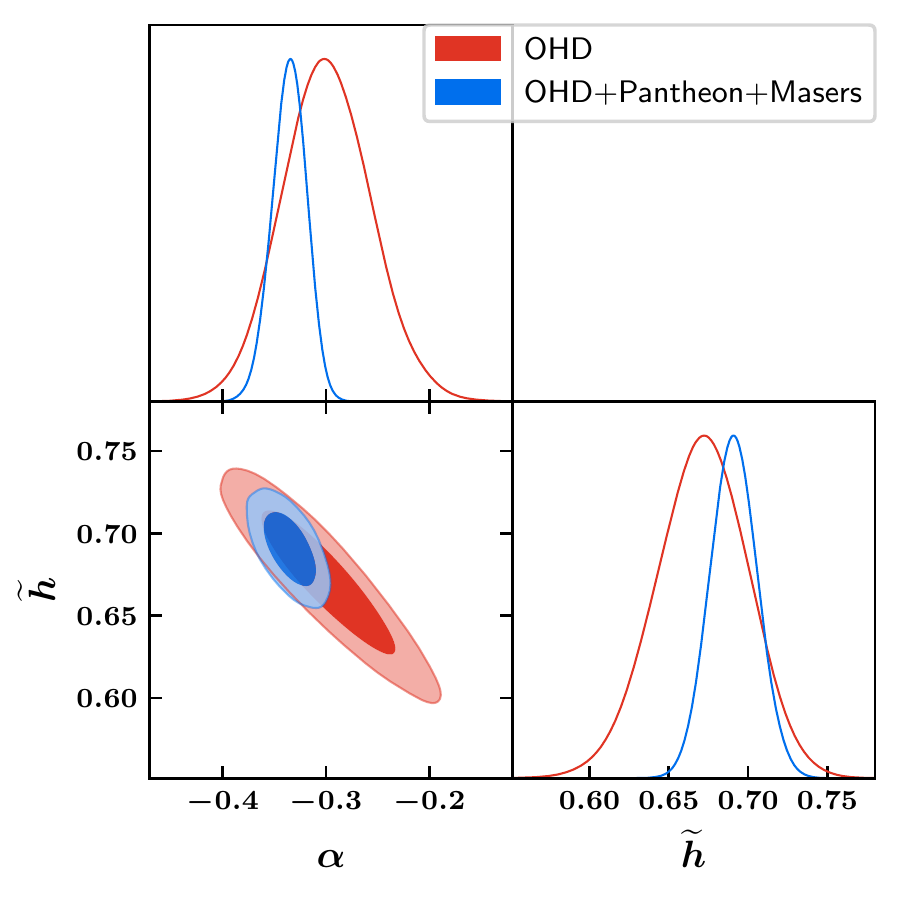}
\caption{Exponential: $2\sigma$ contour levels between $\alpha$ and $\widetilde{h}$ for OHD and its combinations with Pantheon+Masers.}
\label{fig:tri_exp}
\end{figure*}

We also perform the similar analysis for the second parametrization, where in this case we show the  parametric dependence between $\alpha$ and $\widetilde{h}$ for two sets of data in fig. (\ref{fig:tri_exp}) and the resulted constraints are shown in the table (\ref{table1}). Here, the value of the Hubble constant is consistent rather with the $DES+ BAO+ Planck$ combined data in the $1$-$\sigma$ level (see table (2) and Eq. (45) of \cite{planckd}). Thus, there is no reduction of the tension to significant order. This is  expected as in the case of exponential parametrization, there is no phantom crossing (before the present epoch) thus the model mimics the $\Lambda$CDM. The Friedmann equation of the latter can be written as : $\widetilde{H}(\widetilde{z}) = 100 \widetilde{h}\left( \widetilde{\Omega}_M (1+\widetilde{z})^3 + 1-\widetilde{\Omega}_M \right)$ where $\widetilde{\Omega}_M$ is the matter density parameter.

\begin{figure*}[!ht] 
\centering
\begin{subfigure}{0.495\linewidth} \centering 
   \includegraphics[height=5.6cm,width=6.8cm]{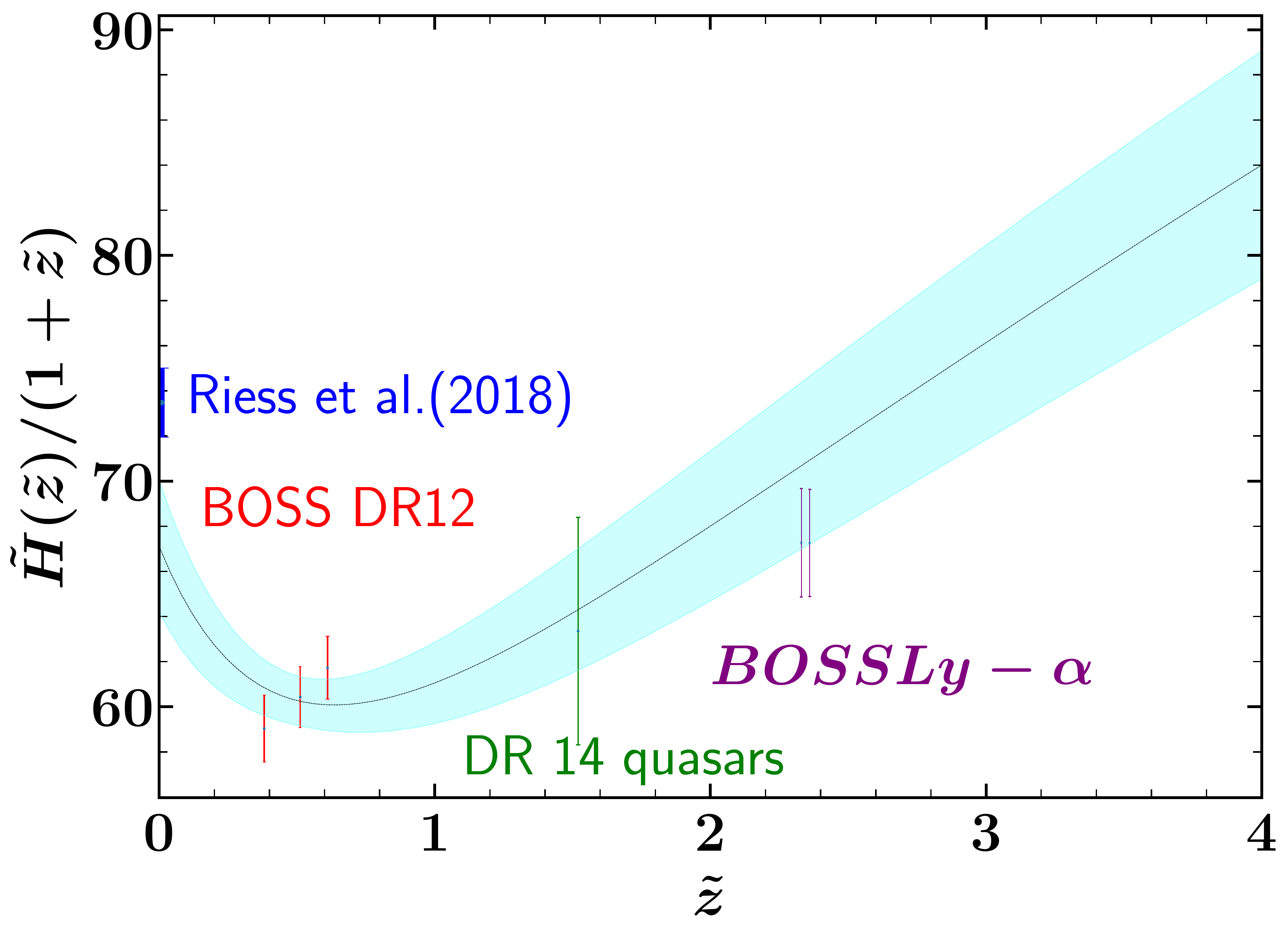}
   \caption{} \label{exp-H-a}
\end{subfigure}
\begin{subfigure}{0.495\linewidth} \centering
    \includegraphics[height=5.6cm,width=6.8cm]{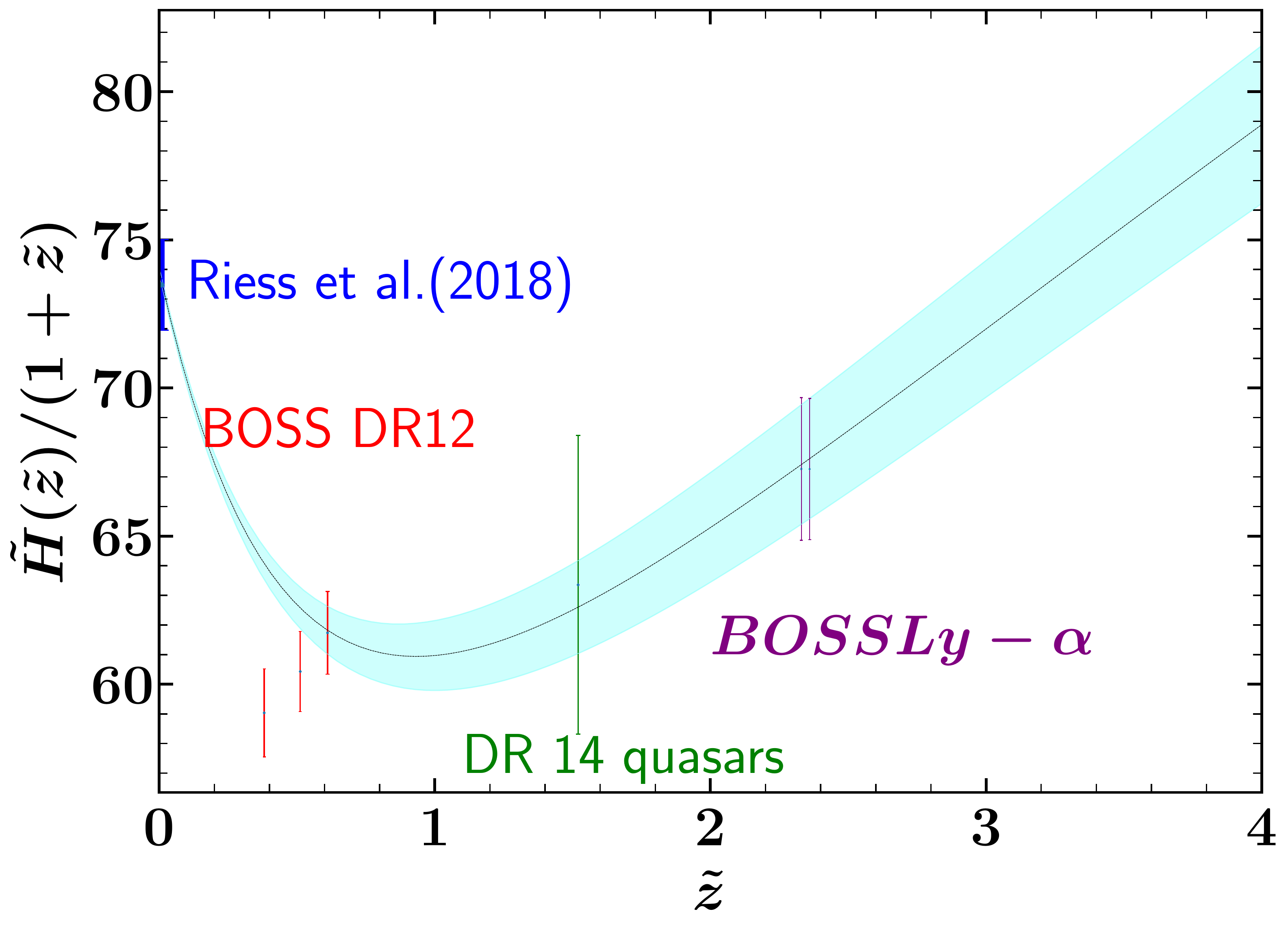}
    \caption{} \label{exp-H-b}
\end{subfigure}
\caption{Exponential: Figures (\ref{exp-H-a}) and (\ref{exp-H-b}) depicts the evolution of $\widetilde{w}_{de}(\widetilde{z})$ with $\widetilde{z} \in [0,4]$ for the datasets OHD and OHD+Pantheon+Masers, respectively. The dark line represents the best-fit and the shaded region corresponds to the $1\sigma$ limit.}
\label{fig:figure3}

\end{figure*} 
\begin{figure*}[!ht] 
\centering
\begin{subfigure}{0.495\linewidth} \centering 
   \includegraphics[height=5.4cm,width=8cm]{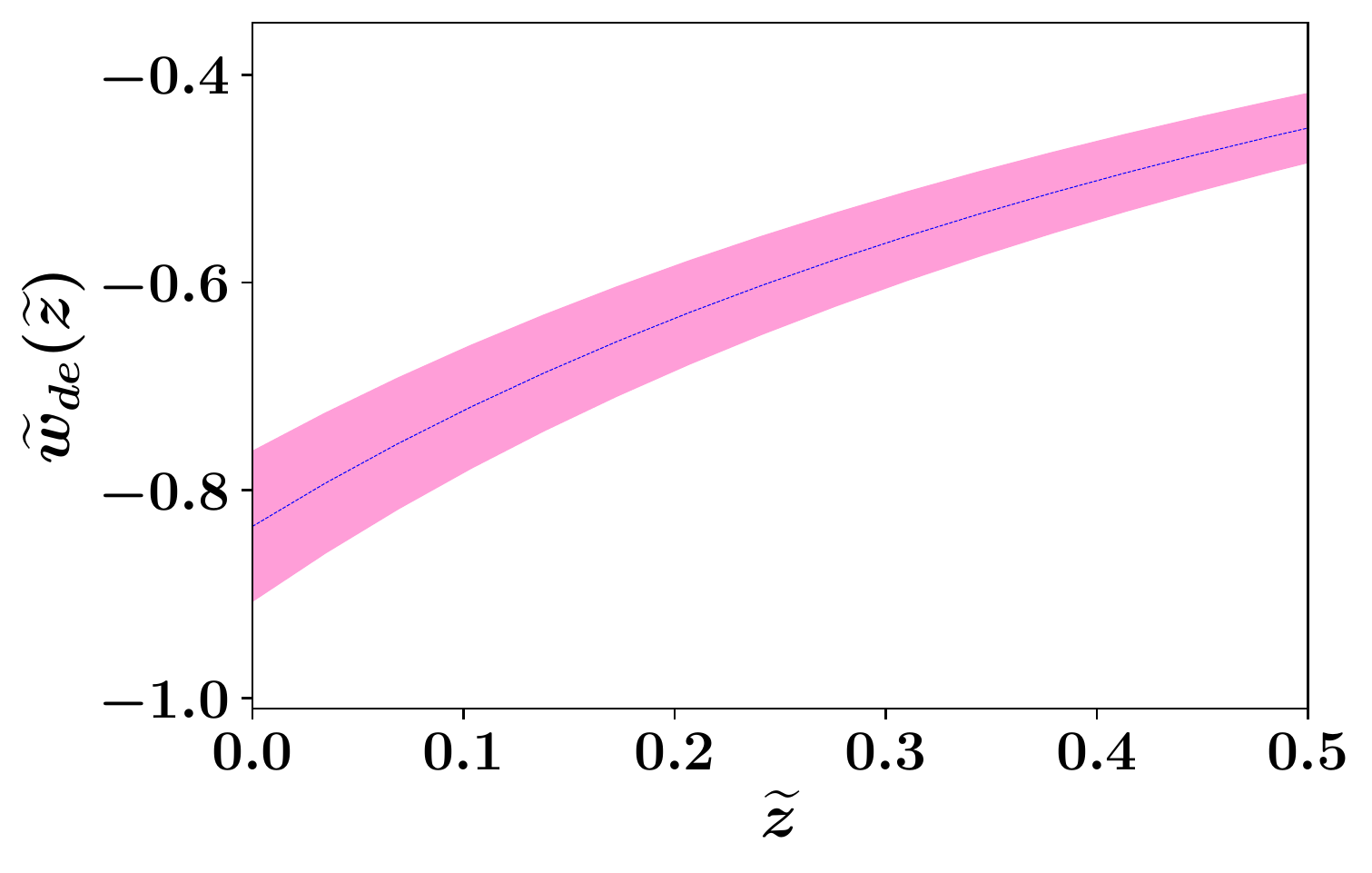}
   \caption{} \label{exp-wde-a}
\end{subfigure}
\begin{subfigure}{0.495\linewidth} \centering
    \includegraphics[height=5.4cm,width=8cm]{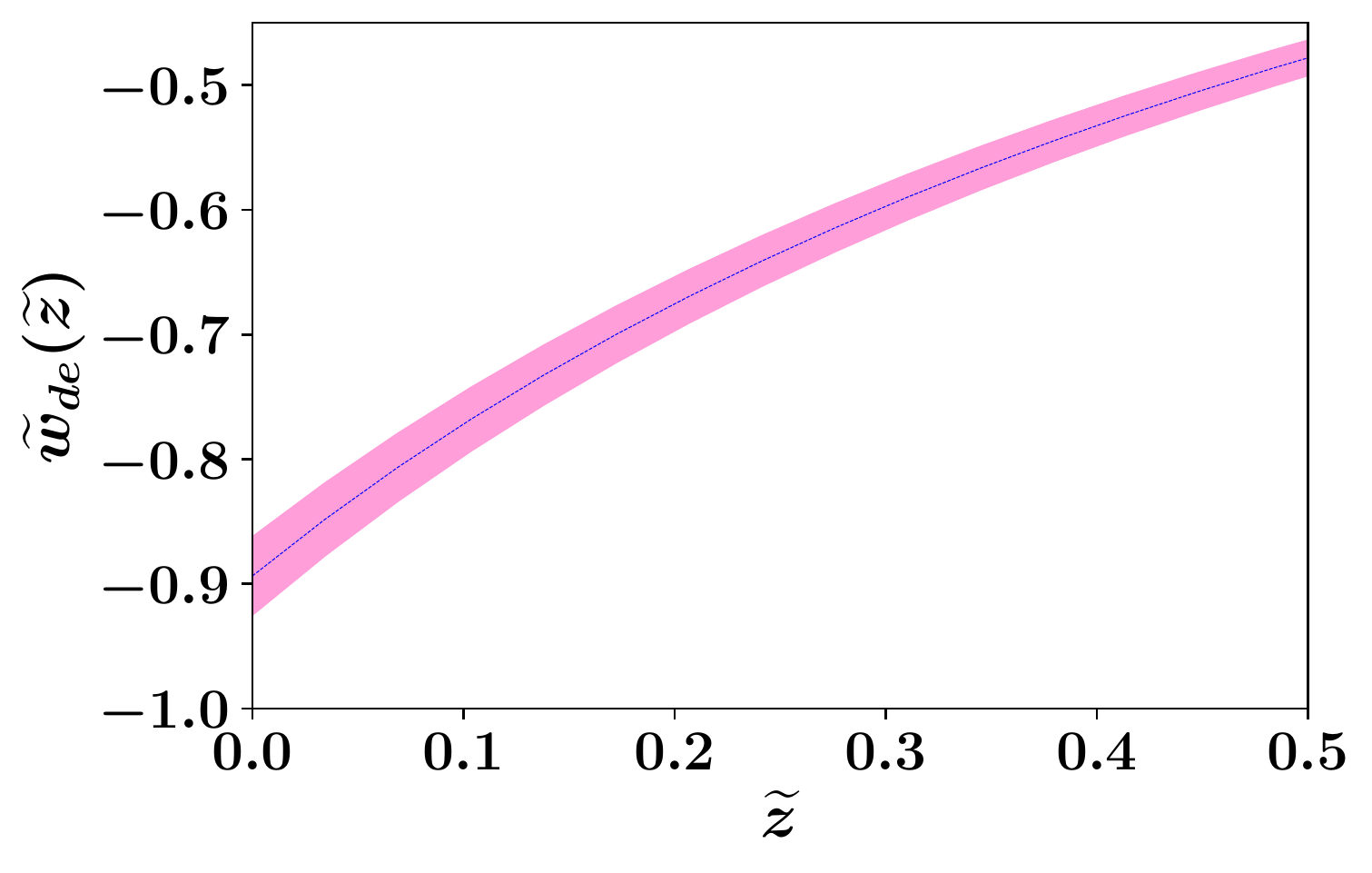}
    \caption{} \label{exp-wde-b}
\end{subfigure}
\caption{Exponential: Figures (\ref{exp-wde-a}) and (\ref{exp-wde-b}) depicts the evolution of $\widetilde{w}_{de}(\widetilde{z})$ with $\widetilde{z} \in [0,4]$ for the datasets OHD and OHD+Pantheon+Masers, respectively. The dark line represents the best-fit and the shaded region corresponds to the $1\sigma$ limit.
}
\label{fig:wde_exp}

\end{figure*}
\begin{figure*}[!ht] 
\centering
\begin{subfigure}{0.495\linewidth} \centering 
   \includegraphics[height=5.5cm,width=8cm]{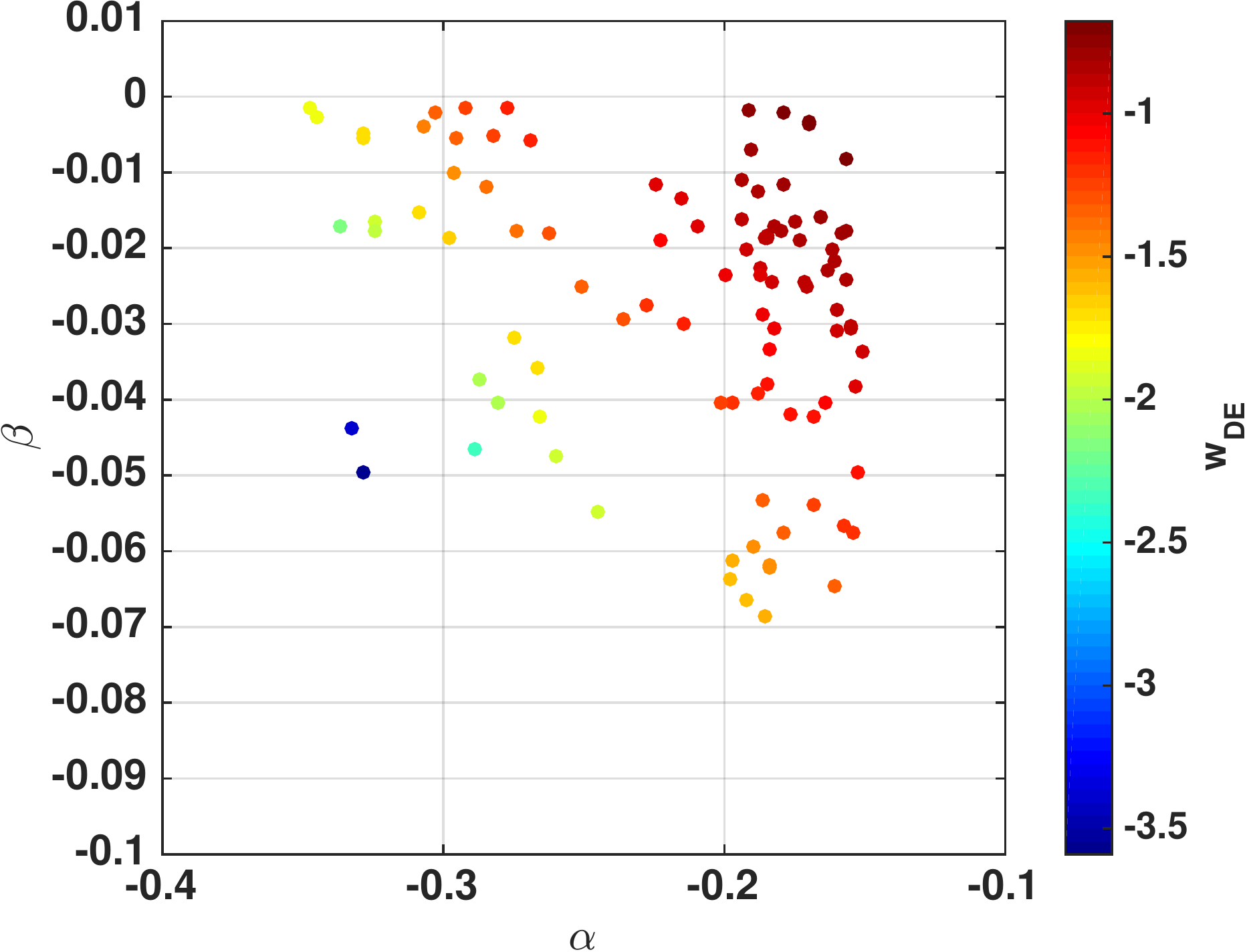}
   \caption{} \label{polyparam}
\end{subfigure}
\begin{subfigure}{0.495\linewidth} \centering
    \includegraphics[height=5.5cm,width=8cm]{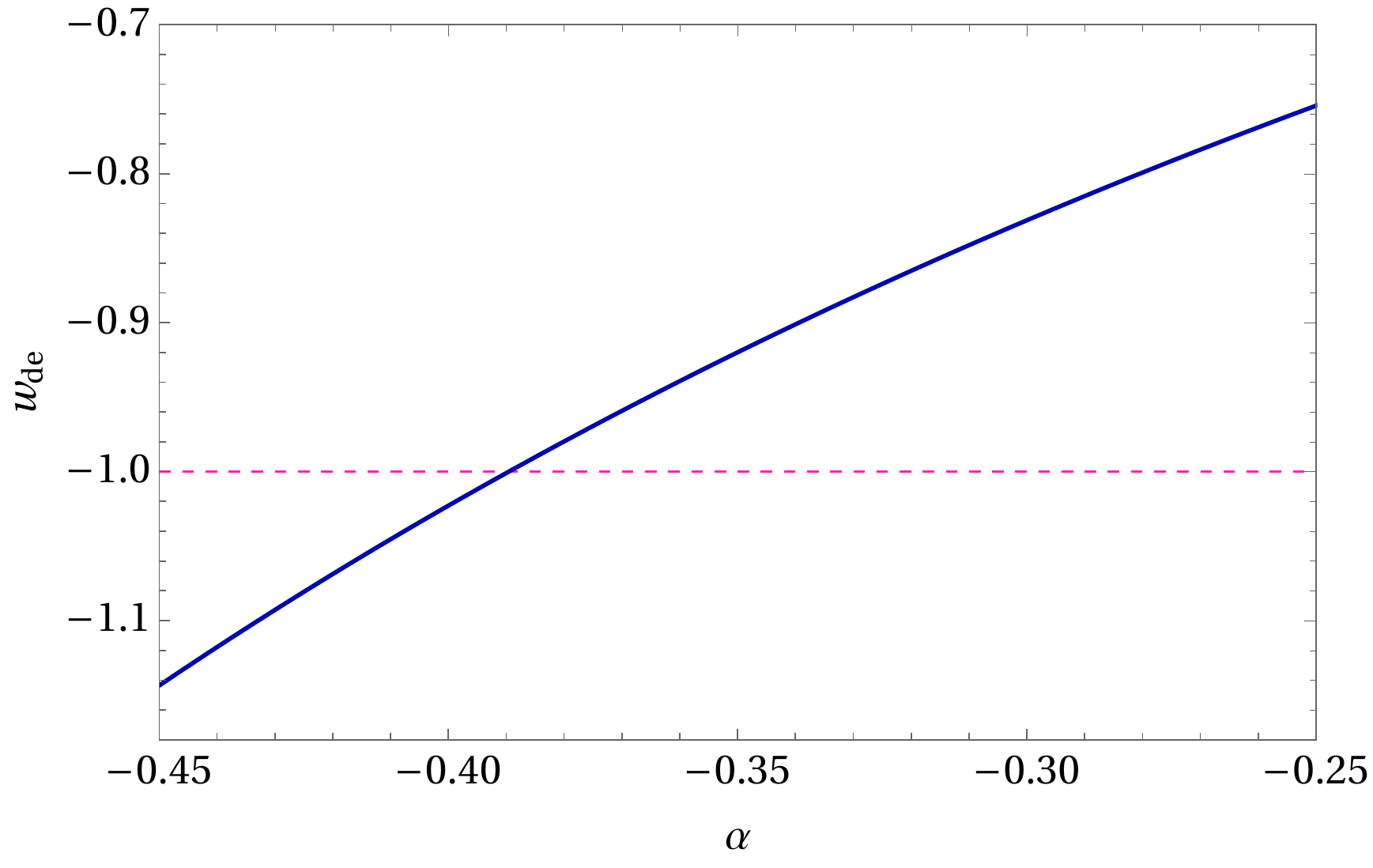}
    \caption{} \label{expparam}
\end{subfigure}
\caption{Exponential: Figures (\ref{polyparam}) and (\ref{expparam}) shows the variation of the two model parameters with equation of state of the dark energy ($w_{DE}$)}
\label{fig:param_wde}

\end{figure*}
Let us note that from the two considered parametrizations, the polynomial case is much better suited to address the Hubble tension. And this can also be understood from the perspective  of statistical significance which is discussed in details in the subsection \ref{biccc}.

From fig. (\ref{fig:figure3}), we see that unlike the polynomial case, the combined data set agrees both with
Riess et. al. and BOSSLy-$\alpha$, which is not the case when using only OHD data set. Moreover, from the fig. (\ref{fig:wde_exp}), one can notice that the phantom crossing of $\widetilde{w}_{de}(\widetilde{z})$ does not happen. In fig. (\ref{fig:param_wde}), we explicitly show the dependence of model parameters on the $\widetilde{w}_{de}$. In fig. (\ref{polyparam}), we show that in order to give obtain the phantom crossing both $\alpha$ and $\beta$ should be negative, similarly, for the exponential case the dependence of $\widetilde{w}_{de}$ on parameter $\alpha$ is shown in fig. (\ref{fig:wde_exp}).
\\
\subsection{Age of  objects in the Universe}

\begin{table*}[!hb]
\centering
\renewcommand{\arraystretch}{1.4}
\begin{tabular}{|c|c||c|}  %
	\hline
	&   \multicolumn{2}{|c|}{Parametrizations} \\
	\cline{2-3}
	Observational &  Polynomial & Exponential \\
	dataset &  Best-fit($\pm 1\sigma$) &  Best-fit($\pm 1\sigma$)\\
	\hline
	 OHD & $\widetilde{t}(0)$ =  $13.657_{-0.421}^{+0.428}$ Gyr & $\widetilde{t}(0)$ = $13.698_{-0.308}^{+0.307} $  Gyr \\
 	 \hline
OHD+Pantheon +Masers	& $\widetilde{t}(0)= 13.917^{+0.366}_{-0.369}$ Gyr  & $\widetilde{t}(0)$ =  $13.526_{-0.237}^{+0.237}$ Gyr\\
	\hline
\end{tabular}
\caption{Age of the Universe with their $1\sigma$ levels for polynomial and exponential parametrizations for OHD and OHD+Pantheon+Masers datasets. }
\label{Table2}
\end{table*}

In case of quintessence, the lower bound on its equation of state parameter $\omega_q= -1$. When effective equation of state reaches this limit, $w_q$ is close to 
$-1$. The late time acceleration is essential for the consistence of hot big bang with observation on the age of universe. Indeed, as acceleration commences ($\tilde{z}\sim 0.5$), the Hubble expansion rate slows down, thereby, it takes more time to reach a particular value of the Hubble parameter, for instance, $\widetilde{H}_0$ and this substantially improves the age of Universe compared to its counter part estimated in absence of late phase of accelerated expansion. In our framework, we have phantom crossing, and compared to the quintessence, in this case, we expect further improvement in age estimate till we cross quintessence lower bound. However, after phantom crossing, the Hubble parameter starts increasing and might match with the one obtained from local observations. But the latter might somewhat suppress the age compared the estimate obtained in the presence of quintessence. Secondly, there are few observations on age of high red-shift objects and since the model under consideration not very different from quintessence or $\Lambda$CDM in the past, it is expected that it would well reconcile with the said data. It is therefore necessary to check, how well, the age of universe in the scenario compares with both the high and low  redshift data, Planck and Globular cluster data, respectively.

\begin{figure*}[!ht] 
\centering
\begin{subfigure}{0.495\linewidth} \centering 
   \includegraphics[height=6.6cm,width=7.5cm]{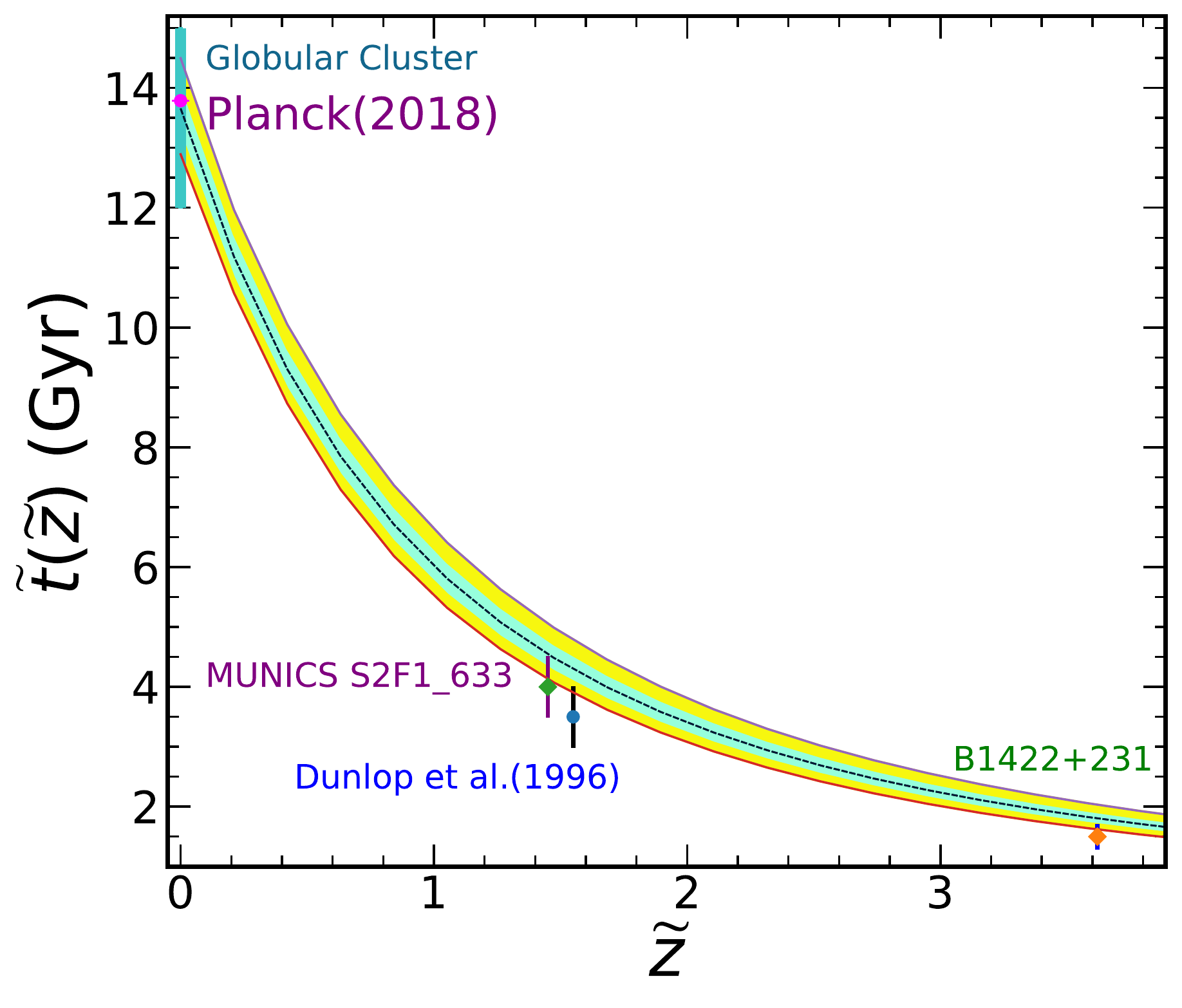}
   \caption{} \label{poly-age-a}
\end{subfigure}
\begin{subfigure}{0.495\linewidth} \centering
    \includegraphics[height=6.6cm,width=7.5cm]{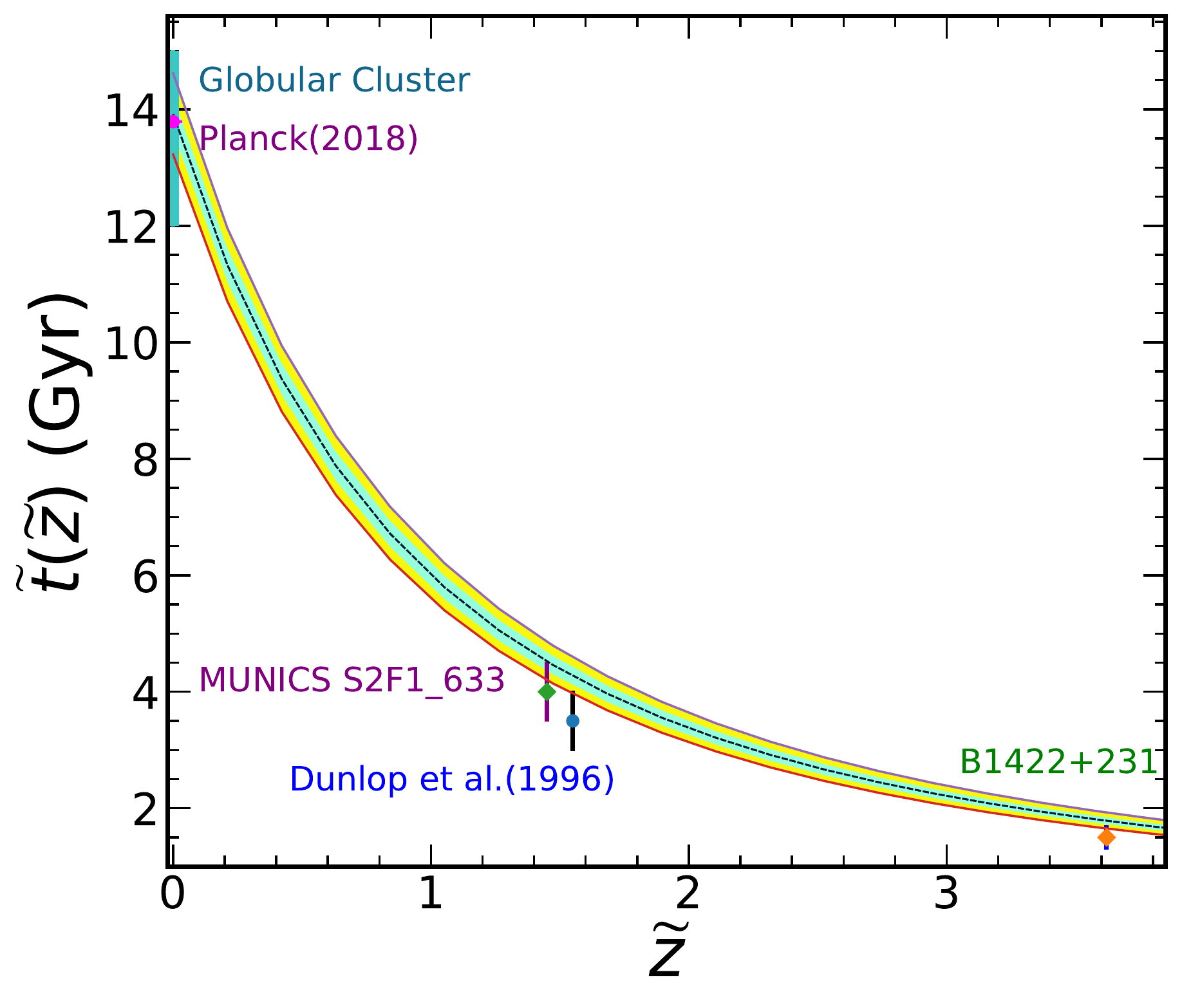}
    \caption{} \label{poly-age-b}
\end{subfigure}
\caption{Polynomial: Figures (\ref{poly-age-a}) and (\ref{poly-age-b}) depict the evolution of $\widetilde{t}_{de}(\widetilde{z})$ with $\widetilde{z} \in [0,4]$ for the datasets OHD and OHD+Pantheon+Masers, respectively. The dark line represents the best-fit, the green and yellow shaded region corresponds to the $1\sigma$ and $2\sigma$ limits respectively.}
\label{fig:poly_age1}
\end{figure*} 

\begin{figure*}[!htb] 
\centering
\begin{subfigure}{0.495\linewidth} \centering 
   \includegraphics[height=6.6cm,width=7.5cm]{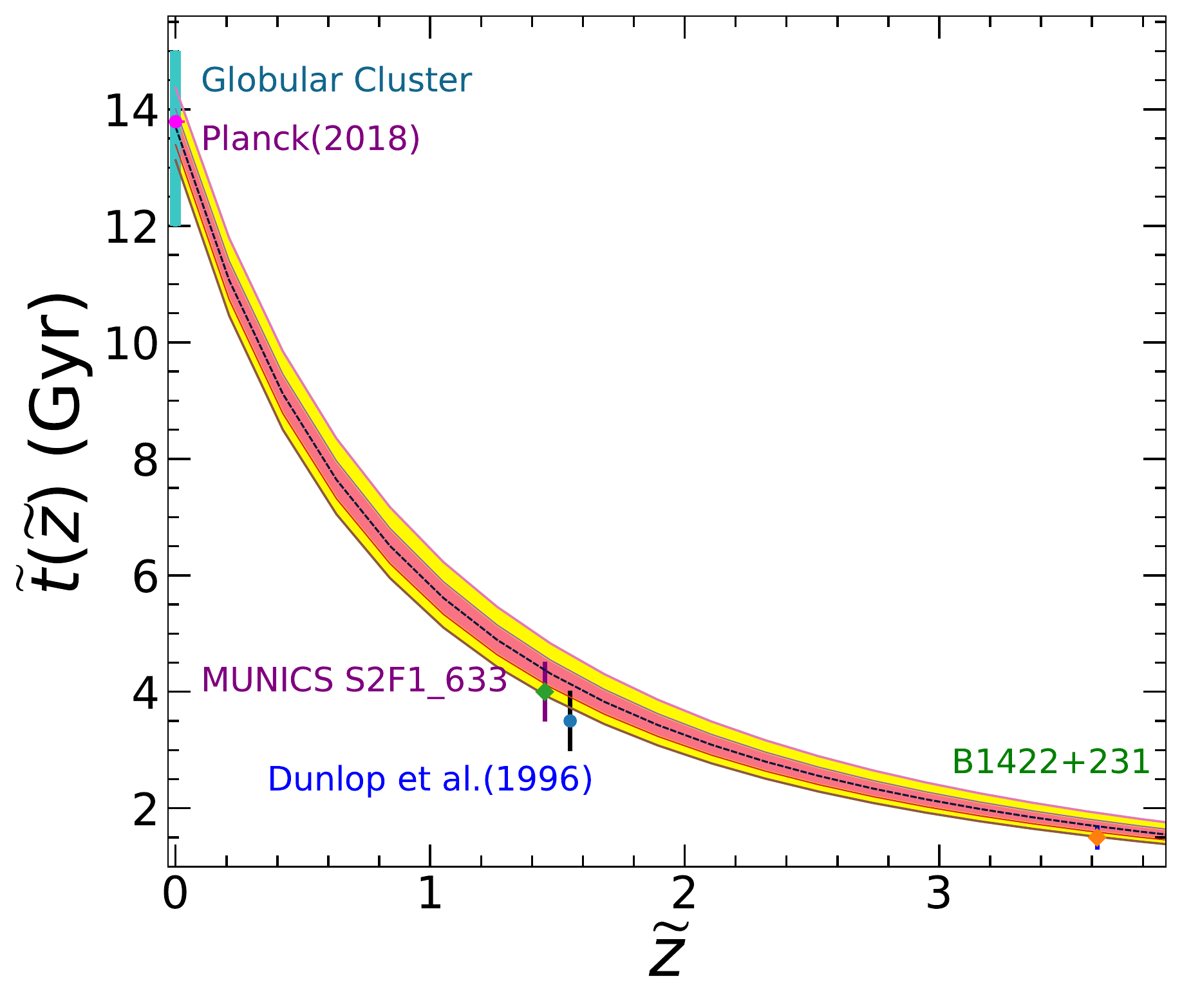}
   \caption{} \label{exp-age-a}
\end{subfigure}
\begin{subfigure}{0.495\linewidth} \centering
    \includegraphics[height=6.6cm,width=7.5cm]{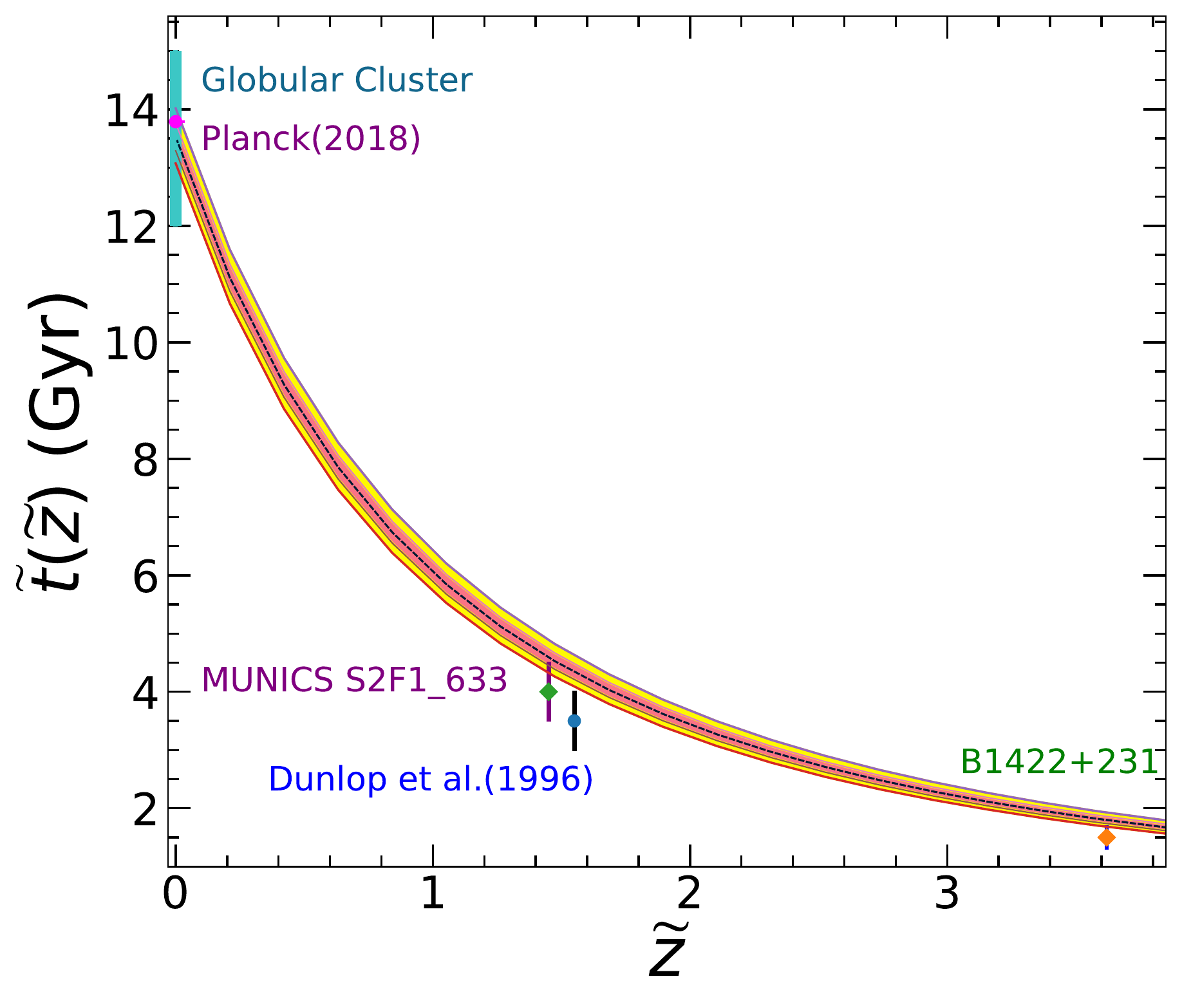}
    \caption{} \label{exp-age-b}
\end{subfigure}
\caption{Exponential: Figures (\ref{exp-age-a}) and (\ref{exp-age-b}) depicts the evolution of $\widetilde{t}_{de}(\widetilde{z})$ with $\widetilde{z} \in [0,4]$ for the datasets OHD and OHD+Pantheon+Masers, respectively. The dark line represents the best-fit , the orange and yellow shaded region corresponds to the $1\sigma$ and $2\sigma$ limits respectively.}
\label{fig:poly_age2}

\end{figure*} 

The age of Universe, for both parametrizations, ($\ref{m1}$) and (\ref{m2}), is given by,
\begin{equation}
\tilde{t}_{(\tilde{z})}=\frac{1}{\widetilde{H}_0}\int_{\tilde{z}}^\infty{\frac{d\tilde{z}}{F(\tilde{z})}}
\end{equation}
where the function $F$ reads as follows,
\begin{equation}
F(\tilde{z})=
\begin{cases}
 & F_{(poly)}(\alpha,\beta ,\tilde{z}) ~~~\text{Polynomial parametrization} \\
&  F_{(exp)}(\tilde{z},\alpha) ~~~\text{Exponential parametrization}
\end{cases}
\end{equation}
where $F_{(poly)}$ and $F_{(exp)}$ are given in Eqs.(\ref{p1-standard}) and ({\ref{a3}}) respectively.

From the obtained constraints, we have found the age of the universe $\widetilde{t}_{de}(\widetilde{z})$ and the corresponding constraints  on $\widetilde{t}_{de}(\widetilde{0})$ as shown in table (\ref{Table2}) for both  parametrizations (\ref{m1}) and (\ref{m2}).
Fig. (\ref{fig:poly_age1}) and (\ref{fig:poly_age2})  show the age of the Universe at a given red-shift for the two paramerizations. These figures also include data points of  
  two high redshift galaxies and the quasar $B1422+231$ along with the globular cluster and Planck'18 results \cite{age,LopezCorredoira2017,age2,Damjanov,age3,LopezCorredoira2018} for comparison. Our results for both the cases are consistent with the 
requirement that the Universe be older than any of its
constituents at a given red-shift.



\subsection{Comparison with $\Lambda$CDM}
\label{biccc}
 To compare the two parametrizations, polynomial (\ref{m1})) and exponential( (\ref{m2})) with the vanilla $\Lambda$CDM model, one needs to take care of the introduction of the extra degrees of freedom with respect to the standard model. In case of polynomial, we have two extra degrees of freedom and in case of exponential, we have one extra degree of freedom. Thus, a detailed Bayesian Information Criterion (BIC) \cite{KassRaftery,DIC} (following \cite{mrg1}) is calculated to take care of that. BIC analysis penalises a model with extra number of degrees of freedom. Thus, to have strong evidence in support of  the model with extra degrees of freedom means much better agreement with observation with respect to the model with less number of degrees of freedom. Under the assumption that the model errors are independent and obey a normal distribution, then the BIC can be rewritten in terms of $\Delta \chi^2$ as $BIC\approx \Delta \chi^2 + df. ln(n)$, where, $df$ is the number of degrees of freedom in the test and $n$ is the number of points in the observed data. The details of our findings are given in the Table (\ref{tablebic}). One can see from Table (\ref{tablebic}), the polynomial parametrization has clearly strong evidence with respect to the standard $\Lambda$CDM scenario. Any evidence where $\Delta BIC \geq 10$ reflects very strong evidence for the new model postulated with respect to the standard one. While, the exponential case has positive evidence in case of the combined data, there is no significant evidence of it when the OHD data is only considered. For the polynomial parametrization, we find strong evidence for both OHD as well as combined data.
\\
\begin{table}[htb]
\centering
\begin{tabular}{|l|l|l|l|l|}
\hline
\begin{tabular}[c]{@{}l@{}}Observational \\      Dataset\end{tabular} & \begin{tabular}[c]{@{}l@{}}Polynomial\\   ($\Delta$ BIC)\end{tabular} & \begin{tabular}[c]{@{}l@{}}Polynomial\\ Evidence\end{tabular} & \begin{tabular}[c]{@{}l@{}}Exponential\\   ($\Delta$ BIC)\end{tabular} & \begin{tabular}[c]{@{}l@{}}Exponential\\   Evidence\end{tabular} \\ \hline
OHD                                                                   & 9.63                                                                   & Strong                                                        & 0.62                                                                  & Not worth                                                       \\ \hline
OHD+Pantheon +Masers                                                  & 8.88                                                                   & Strong                                                        & 4.01                                                                  & Positive                                                        \\ \hline
\end{tabular}
\caption{The evidence in support of polynomial and exponential parametrizations for OHD and OHD+Pantheon+Masers datasets with respect to the standard $\Lambda$CDM scenario. }
\label{tablebic}
\end{table}

\FloatBarrier

\section{Conclusion and future perspectives }
\label{conclusion}

In this paper, we have investigated the observational viability and generic implications of the dark matter  and baryonic matter  interaction in the Einstein frame, which is caused by a general disformal transformation between the Jordan and the Einstein frames. In particular, the phenomenon is based upon the assumption that dark matter follows the Einstein frame geodesics, whereas, the baryonic matter obeys Jordan frame trajectories. Consequently, under the standard disformal transformation, a coupling is induced between both the matter components  which spoils their individual  energy conservation in the Einstein frame. 

As the geodesics of both frames are not equivalent (due to the disformal transformation between them), we choose two different parametrizations to relate the scale factors of both the frames in the standard FRW space-time. In particular, we resort to the polynomial and exponential parametrizations and find the constraints on the model parameters in the Jordan frame considering it to be the physical frame (as the underline mechanism assumes the baryonic matter to follow its trajectories in the Jordan frame and hence all the observations are being done in this frame). In case of the polynomial parametrization Eqn. (\ref{m1}), the best-fit values of Hubble parameter for two different data combinations is  such that it  significantly reduces the so called `Hubble tension'. For OHD data, the tension is insignificant in case of polynomial parametrization, as the computed value of $\widetilde{h}= 0.7279^{+0.05}_{-0.05}$ $1$-$\sigma$ consistent with Riess et al., whereas  for the combined OHD+Pantheon+Masers data, the tension is reduced to $1.3$-$\sigma$ level. We would like to once again emphasise that this is related to the fact that, in this particular parametrization, the dark energy equation of state crosses from  quintessence to phantom region. Hence, we see a strong evidence in support of this parametrization while performing the $\Delta$BIC analysis which is quoted in table (\ref{tablebic}). 

As for the exponential parametrization (Eqn. \ref{m2}), we have not found any significant reduction of Hubble-tension (see, fig. (\ref{fig:tri_exp})) with both the data combinations and this can be attributed to the fact that, the model exhibits quintessence like behavior with $\widetilde{w}_{de}(\widetilde{z})\geq -1$ around the present epoch. Thus as expected, in the  exponential case, we only find some positive evidence in combined data scenario but not significant enough when only OHD is considered (see table (\ref{tablebic})). 
However, the desired behaviour is achieved in case of the polynomial parametrization where  phantom-crossing takes place. To this effect, a universal Bayesian evidence calculation could robustly support our claim.

In the framework under consideration, the behavior of the Hubble parameter at late times is generically different from quintessence which is also reflected on the estimate of age of Universe. While for checking the consistency with the globular clusters, Planck 2018 results and the high red-shift data, we find an excellent agreement
  with observations for both the parametrizations (see figs. (\ref{poly-age-a}, \ref{poly-age-b}) and (\ref{exp-age-a}, \ref{exp-age-b})).
  
It is worth  emphasizing that the scenario  based upon interaction between DM and BM  admits late-time cosmic acceleration without invoking any exotic fluid and it is compatible with observation. One of the most important and generic  implications of the  interaction at the background level 
includes the presence of phantom-crossing  which is supported by most of the observations at present. 

 It will be interesting to address the issues like Hubble tension using the different dynamical realisation of the early time acceleration , say,  `warm' inflation \cite{mar1, sharma} in this framework.  Also  in case of non-canonical realisation of inflation \cite{Bhattacharya}, one observes non standard sound speed ($c_s$) which then can have implications on the measured value of $\widetilde{H}_0$ from CMB. The baryon asymmetry of the Universe can be studied also in this framework based on the ideas proposed in \cite{novikov1, novikov2, novikov3, novikov4}  the In our opinion, this is an interesting investigation  to be carried out within the framework of the  model  considered here.
 Last but not least, it would be interesting to consider  perturbations
 and study matter power spectrum in the presence of disformal coupling between DM and BM  following Refs. \cite{Bhattacharya:2020zap1,Bhattacharya:2020zap2}. We would like to address all these issues in our upcoming endeavors. 
 \\
\\
\textbf {Acknowledgments:} The authors would like to thank Akash Garg, Nur Jaman, Sibesh K. Jas Pacif, Ruchika and Somasri Sen for useful discussions. Work of MRG is supported by Department of Science and Technology(DST), Government of India under the Grant Agreement number IF18-PH-228 (INSPIRE Faculty Award) and by Science and Engineering Research Board(SERB), Department of Science and Technology(DST), Government of India under the Grant Agreement number CRG/2020/004347(Core Research Grant). MS is supported by the Ministry of Education and Science of the Republic of Kazakhstan, Grant
No. 0118RK00935 and by NASI-Senior Scientist Platinum Jubilee Fellowship(2021). The work of MKS is supported by the Council of Scientific and Industrial Research (CSIR), Government of India. The authors would also like to thank the anonymous referee for valuable suggestions. 
\\
\section{Appendix}
\subsection{Fractional density parameters} \label{app-p1-dm}
As we have acceleration in the Jordan frame by virtue of a disformal coupling, it would be convenient to express (20) in the conventional form by isolating   
  the term proportional to $(1+\widetilde{z})^{3}$  $\hat{\rm a}$  {\it  la} the effective fractional density of cold matter and remaining terms
  can be assigned the role of 
   dark energy.

\begin{equation}
\frac{\widetilde{H}^{2}}{\widetilde{H}_{0}^{2}}=A(\alpha ,\beta )(1+\widetilde{z}%
)^{3}+A(\alpha ,\beta )f(\widetilde{z})\text{,}  \label{34}
\end{equation}%
where  
\begin{eqnarray}
A(\alpha ,\beta ) &=&(1+\alpha +\beta )(1+2\alpha +3\beta )^{2}\text{,}
\label{34A} \\
f(\widetilde{z}) &=&-5(1+\widetilde{z}^{2})\alpha -\alpha \left( 49\alpha
^{2}-48\beta \right) +(1+\widetilde{z})\left( 17\alpha ^{2}-7\beta \right) 
\notag \\
&&+\frac{(1+\widetilde{z})\alpha ^{6}-5(1+\widetilde{z})\alpha ^{4}\beta +\alpha
^{5}\beta +6(1+\widetilde{z})\alpha ^{2}\beta ^{2}-4\alpha ^{3}\beta ^{2}-(1+%
\widetilde{z})\beta ^{3}+36\alpha \beta ^{3}}{\left( \alpha ^{2}-4\beta \right)
\left( (1+\widetilde{z})^{2}+(1+\widetilde{z})\alpha +\beta \right) }  \notag \\
&&+\frac{%
\begin{array}{c}
128(1+\widetilde{z})\alpha ^{6}-64\alpha ^{7}-720(1+\widetilde{z})\alpha ^{4}\beta
+576\alpha ^{5}\beta +864(1+\widetilde{z})\alpha ^{2}\beta ^{2} \\ 
-1512\alpha ^{3}\beta ^{2}-135(1+\widetilde{z})\beta ^{3}+918\alpha \beta ^{3}%
\end{array}%
}{\left( \alpha ^{2}-4\beta \right) \left( (1+\widetilde{z})^{2}+2\alpha (1+%
\widetilde{z})+3\beta \right) }  \notag \\
&&+\frac{%
\begin{array}{c}
128(1+\widetilde{z})\alpha ^{8}-960(1+\widetilde{z})\alpha ^{6}\beta +192\alpha
^{7}\beta +2160(1+\widetilde{z})\alpha ^{4}\beta ^{2}-1296\alpha ^{5}\beta ^{2}
\\ 
-1512(1+\widetilde{z})\alpha ^{2}\beta ^{3}+2376\alpha ^{3}\beta ^{3}+162(1+%
\widetilde{z})\beta ^{4}-1053\alpha \beta ^{4}%
\end{array}%
}{\left( \alpha ^{2}-4\beta \right) \left( (1+\widetilde{z})^{2}+2\alpha (1+%
\widetilde{z})+3\beta \right) ^{2}}\text{,}  \label{35}
\end{eqnarray}%

The  Friedmann equation in Jordan-frame can be cast in terms of fractional energy densities,
\begin{equation}
{\widetilde{H}^{2}}=\widetilde{H}_{0}^{2}\left[ \Omega _{Meff}^{\left( 0\right) }(1+%
\widetilde{z})^{3}+\Omega _{DE}^{\left( 0\right) }F(\widetilde{z})\right] \text{,}
\label{36}
\end{equation}%
where, $\Omega _{Meff}^{\left( 0\right) }\equiv A$ and $\Omega
_{DE}^{\left( 0\right) }\equiv Af(0)$ 
and $F(\widetilde{z})\equiv {f(\widetilde{z})}/{f(0)}$. 

\end{document}